\documentclass[onecolumn,journal,letterpaper]{IEEEtran}

 \usepackage{color}
\usepackage{amsmath}
\usepackage{theorem}
\usepackage{dsfont}
\usepackage{amsfonts}
\usepackage{amssymb}
\usepackage{cite}
\usepackage{cancel}
\usepackage{dsfont}

\usepackage[usenames,dvipsnames]{pstricks}
\usepackage{epsfig}

\usepackage{pst-poly}
\usepackage{pst-node}
\usepackage{pst-plot}
\usepackage{pst-grad}
\newtheorem{thm}{Theorem}

\newtheorem{coro}{Corollary}

\newcommand{\SNR}{{\sf SNR}}
\newcommand{\DoF}{{\sf DoF}}

\newcommand{\T}{\mathcal{T}}
\newcommand{\vR}{{\bf R}}

\newcommand{\td}{\tilde{d}}

\begin{document}
\thispagestyle{empty}

\pagestyle{empty}

\title{Completely Stale Transmitter Channel State Information is Still Very Useful\thanks{This research is partially supported by a gift from Qualcomm Inc. and by the AFOSR under grant number FA9550-09-1-0317.
}\thanks{
An initial version of this paper has been reported as Technical Report No. UCB/EECS-2010-122 at the University of California--Berkeley, Sept. 6, 2010.}
\thanks{This paper has been partially presented in the Forty-Eighth Annual Allerton Conference on Communication, Control, and Computing, Allerton Retreat Center, Monticello, Illinois, Sept. 2010.}
\thanks{Mohammad A. Maddah-Ali is currently with Bell-Labs Alcatel--Lucent. This work was done when he was with the University of California--Berkeley as a post--doctoral fellow.}
\thanks{Copyright (c) 2011 IEEE. Personal use of this material is permitted.  However, permission to use this material for any other purposes must be obtained from the IEEE by sending a request to pubs-permissions@ieee.org.}
}

\author{Mohammad Ali Maddah-Ali and David Tse\\
Wireless Foundations,\\
Department of Electrical Engineering and Computer Sciences,\\
University of California, Berkeley
}
 \maketitle

\thispagestyle{empty}

\pagestyle{empty}

\begin{abstract}
Transmitter channel state information (CSIT) is crucial for the multiplexing gains offered by advanced interference management techniques such as multiuser MIMO and interference alignment. Such CSIT is usually obtained by feedback from the receivers, but  the feedback is subject to delays. The usual approach is to use the fed back information to predict the current channel state  and then apply a scheme designed assuming perfect CSIT. When the feedback delay is large compared to the channel coherence time, such a prediction approach completely fails to achieve any multiplexing gain. In this paper, we show that even in this case, the completely stale CSI is still very useful. More concretely, we show that in a MIMO broadcast channel with $K$ transmit antennas and $K$ receivers each with $1$ receive antenna,  $\frac{K}{1+\frac{1}{2}+ \ldots+ \frac{1}{K}} (> 1) $ degrees of freedom is achievable even when the fed back channel state  is completely independent of the current channel state. Moreover, we establish that if all receivers have independent and identically distributed channels, then this is the optimal number of degrees of freedom achievable. In the optimal scheme, the transmitter uses the fed back CSI to learn the side information that the receivers receive from previous transmissions rather than to predict the current channel state. Our result can be viewed as the first example of feedback providing a degree-of-freedom gain in memoryless channels.
\end{abstract}
%
%

\section{Introduction}

In wireless communication, transmitter knowledge of the channel
state information (CSIT) can be very important. While in
point-to-point channels CSIT only provides power gains via
waterfilling, in multiuser channels it can also provide {\em
multiplexing gains}. For example, in a MIMO broadcast channel,  CSIT
can be used to send information along multiple beams  to different
receivers simultaneously. In interference channels, CSIT can be used
to align the interference from multiple receivers to reduce the
aggregate interference footprint
\cite{maddah_x_journal,cadambe2008iaa}.

In practice, it is not easy to achieve the theoretical gains of these techniques. In the high $\SNR$ regime, where the multiplexing gain offered by these techniques is particularly significant, the performance of these techniques is very sensitive to inaccuracies of the CSIT. However, it is hard to obtain accurate CSIT. This is particularly so in FDD (frequency-division duplex) systems, where the channel state has to be measured at the receiver and fed back to the transmitter. This feedback process leads to two sources of inaccuracies:

\begin{itemize}
\item \emph{Quantization Error:} The limited rate of the feedback channel restricts the accuracy of  the CSI at the transmitter.

\item \emph{Delay:} There is a delay between the time the channel state is measured at the receiver and the time when the information is used at the transmitter. The delay comes from the fact that the receivers need some time to receive pilots, estimate CSI, and then feed it back to the transmitter in a relatively long coding block. In time-varying wireless channels, when the channel information arrives at the transmitter, the channel state has already changed.
\end{itemize}

Much work in the literature has focused on the first issue. The general conclusion is that the rate of the feedback channel needed to achieve the perfect CSIT multiplexing gain scales well with the $\SNR$.  For example, for the MIMO broadcast channel, it was shown in~\cite{Jindal_Finite_Feedback} that the rate of feedback should scale linearly with $\log_2 \SNR$. Since the capacity of the MIMO broadcast channel also scales linearly with $\log_2 \SNR$, this result says that the overhead from feedback will not overwhelm the capacity gains.

 We now focus on the second issue, the issue of feedback delay.
The standard approach of dealing with feedback delay is to exploit
the time correlation of the channel to predict the current channel
state from the delayed measurements~\cite{Caire_Delay}. The predicted channel state is
then used in place of the true channel state in a scheme designed
assuming perfect CSIT is available. However, as the coherence time
of the channel becomes shorter compared to the feedback delay, due
to higher mobility for example, the  delayed feedback information
reveals no information about the current state, and a
prediction-based scheme can offer no multiplexing gain.

In this paper, we raise the question: is this a fundamental
limitation imposed by feedback delay, or is this just a limitation
of the prediction-based approach? In other words, is there another
way to use the delayed feedback information to achieve non-trivial
multiplexing gains? We answer the question in the affirmative.

For concreteness, we focus on a channel which has received
significant attention in recent years: the MIMO broadcast channel.
In particular, we focus on a  system where the transmitter
has $M$ antennas and there are $K$ receivers each with a single
receive antenna. The transmitter wants to send an independent data
stream to each receiver. To model completely outdated CSI, we allow
the channel state to be independent from one symbol time to the
next, and the channel state information is available to both the
transmitter and the receivers {\em one symbol time later}. This
means that by the time the feedback reaches the transmitter, the
current channel is already {\em completely} different. We also
assume that the overall $M$-by-$K$ channel matrix is full rank at
each time.

Our main result is that, for $M \ge  K$, one can achieve a total of
$$\frac{K}{1+\frac{1}{2}+ \ldots+ \frac{1}{K}}$$ degrees of freedom per second per Hz in this channel. In other words, we can achieve a sum rate that scales like:
$$\frac{K}{1+\frac{1}{2}+ \ldots+ \frac{1}{K}} \log_2 \SNR + o(\log_2 \SNR) \quad \mbox{bits/s/Hz}$$
as the $\SNR$ grows. Moreover, we show that under the further
assumption that all receivers have independent and identically distributed channels,
this is the optimal number of degrees of freedom achievable.

It is instructive to compare this result with the case when there is
no CSIT and the case when there is perfect CSIT. While the capacity
or even the number of degrees of freedom is unknown for general
channel statistics when there is no CSIT, in the case when all
receivers have identically distributed channels, it is easy to see
that the total number of degrees of freedom is only $1$. Since
$\frac{K}{1+\frac{1}{2}+ \ldots+ \frac{1}{K}} > 1$ for any $K \ge
2$, we see that, at least in that case, there is a multiplexing gain
achieved by exploiting completely outdated CSI. However, the
multiplexing gain is not as good as $K$, the number of degrees of
freedom achieved in the perfect CSIT case. On the other hand, when
$K$ is large,
$$ \frac{K}{1+\frac{1}{2}+ \ldots+ \frac{1}{K}} \approx \frac{K}{\ln K},$$
almost linear in $K$.

Why is outdated CSIT useful? When there is perfect CSIT, information
intended for a receiver can be transmitted to that receiver without
other receivers overhearing it (say by using a zero-forcing
precoder), so that there is no cross-interference. When the
transmitter does not know the current channel state, this cannot be
done and information intended for a receiver will be overheard by
other receivers. This overheard {\em side information} is in the
form of a  linear combination of data symbols, the coefficients of
which are the channel gains at the time of the transmission. Without
CSIT at all, this side information will be wasted since the
transmitter does not know what the coefficients are and hence does
not know what side information was received in previous
transmissions. With outdated CSIT, however,  the transmitter can
exploit the side information already received at the various
receivers to create future transmissions which are simultaneously
useful for more than one receiver and can therefore be efficiently
transmitted. Note that there is no such overheard side-information
in simpler scenarios such as point-to-point and multiple access
channels, where there is only a single receiver. Indeed, it is shown
in \cite{Vis_delay,BSP11} that for such channels, the only role of
delayed CSIT is to predict the current state, and when the delayed
CSIT is independent of the current state, the delayed CSIT provides
no capacity gains.

The rest of the paper is structured as follows. In Section
\ref{sec:problem_formulation}, the problem is formulated and the
main results are stated precisely. Sections \ref{sec:achievable-square}, \ref{Sec:non-square:original}, and~\ref{Sec:Imoroved_2Tr} describe the proposed schemes, and
Section \ref{sec:outer} describes the converse. In Section~\ref{Sec:DoF-Region}, the ${\DoF}$ region for the case of $M=K$ is characterized. 
The connection
between our results and those for the packet erasure broadcast
channel is explained in Section \ref{sec:bec}. Some follow-up results
to the conference version of this paper are discussed in
\ref{sec:followup}. We conclude with a discussion of our result in
the broader context of the role of feedback in communication in
Section \ref{sec:concl}.

\section{Problem Formulation and Main Results}\label{sec:problem_formulation}
We consider a complex baseband broadcast channel with  $M$ transmit
antennas and $K$ receivers, each equipped with a single antenna. In
a flat fading environment, this channel can be modeled as,
\begin{eqnarray}\label{eq:channel_model}
y_r[n]={\mathbf{h}}_{r}^{\dagger} [n]\mathbf{x}[n]+z_r[n], \quad
r=1,\ldots,K,
\end{eqnarray}
where ${\dagger}$ denotes transpose--conjugate operation, $\mathbf{x}[n] \in \mathbb{C}^{M \times 1}$,
$\mathbb{E}[\mathbf{x}^{\dagger} [n] \mathbf{x}[n]]\leq \SNR$,  $z_r[n]
\sim \mathcal{CN}(0,1)$ and the sequences $z_r[n]$'s are i.i.d. and
mutually independent. In addition, $\mathbf{h}^{\dagger}_{r}[n] =
[h^{\dagger}_{r1}[n],\ldots, h^{\dagger}_{rM}[n] ] \in \mathbb{C}^{1 \times M}$. We
define $\mathbf{H}[n]$ as  $\mathbf{H}[n]=[ \mathbf{h}_{1}[n],
\ldots, \mathbf{h}_{K}[n]]$.

We assume that  $\mathbf{H}[n]$ is available at the transmitter and
all receivers with one unit delay\footnote{All our achievable
results hold regardless of what the delay is, since they do not
depend on the temporal statistics of the channel. Hence, for
convenience, we will just normalize the delay to be $1$ symbol
time.}

Let us define $\mathcal{E}$ as  $\mathcal{E} = \{1,2,\ldots,K\}$. We assume that for any subset  $\mathcal{S}$ of the receivers, $\mathcal{S} \subset \mathcal{E}$, the transmitter has a message $W_{\mathcal{S}}$ with rate $R_{\mathcal{S}}$ bits/s/Hz. For example, message $W_{\{1,2\}}$ is a common message for receivers one and two. Similarly, $W_{\{1\}}$, or simply $W_{1}$, is a message for receiver one.   We define $d_{\mathcal{S}}$, as
\begin{align}
d_{\mathcal{S}} = \lim_{\SNR \rightarrow  \infty }
\frac{R_{\mathcal{S}}}{\log_2 \SNR}.
\end{align}

If $|\mathcal{S}|=j$, then we call $W_{\mathcal{S}}$  an order--$j$
message or a message of order $j$. We define degrees of freedom
order $j$, ${\DoF}^*_j(M,K)$, as
\begin{align}
{\DoF}^*_j (M,K)= \lim_{\SNR \rightarrow  \infty } \max_{\vR \in
\mathcal{C}} \sum_{S, |S| = j } \frac{R_{\mathcal{S}}}{\log_2 \SNR}.
\end{align}
where $\mathcal{C}$ denotes the capacity region of the channel, and
$\vR \in \mathbb{R}^{(2^{K}-1) \times 1}$ denotes the vector of the
message rates for each subset of receivers.
We note that ${\DoF}^*_1 (M,K)$ is the well-known notion of the
degrees of freedom  of the channel.

In this paper, we establish the following results.

\begin{thm} \label{thm:squre}
As long as $\mathbf{H}[n]$ is full rank almost surely for each $n$,
and $\{\mathbf{H}[n]\}$ is stationary and ergodic, then for $M \ge
K$,
\begin{align}
{\DoF}^*_1(M, K) \ge \frac{K}{1+\frac{1}{2}+\ldots+\frac{1}{K}}.
\end{align}
More generally, as long as $M \geq K-j+1$, then
\begin{align}
\label{eq:lower}
 {\DoF}^*_j(M, K) \ge \frac{K-j+1}{j} \frac{1}{
\frac{1}{j}+\frac{1}{j+1}+\ldots+\frac{1}{K} }.
\end{align}
\end{thm}
For example, ${\DoF}^*_1(2,2) \ge \frac{4}{3}$ and ${\DoF}^*_1(3,3) \ge
\frac{18}{11}$, which are greater than one. Note that this
achievability result holds under very weak assumptions about the
channel statistics.  Hence, even when $\{\mathbf{H}[n]\}$ is an
i.i.d.   process over time, delayed CSIT is still useful in achieving
a degree-of-freedom gain.

The following theorem gives a tight converse under specific
assumptions on the channel process.

\begin{thm}\label{thm:outer--bound--Gen}
If the channel matrices $\{\mathbf{H}[n]\}$ is an i.i.d. process over
time and the channels are also independent and identically
distributed across the receivers, then
\begin{align}
\label{eq:upper}
 {\DoF}^*_j(M,K) \leq \frac{    {K \choose j}
}{\frac{ { K-1 \choose j-1}}{ \min \{1, M\} } +\frac{{ K-2 \choose
j-1}}{ \min\{2, M \}}+\ldots + \frac{ {j-1  \choose j-1} }{
\min\{K-j+1, M\} }}
\end{align}
\end{thm}

The equality between the expressions in (\ref{eq:lower}) and
(\ref{eq:upper}) in the case of $M \ge K-j+1$ can be verified using
the identity~\eqref{eq:identity}, proved in
Appendix~\ref{Appendix-A}, thus yielding the following corollary.

\begin{coro}
\label{coro:outer--bound} If the channel matrices
$\{\mathbf{H}[n]\}$ is an i.i.d.  process over time and is also
independent and identically distributed across the receivers, then
the lower bounds in Theorem \ref{thm:squre} are tight.
\end{coro}

In addition, the region of order--one $\DoF$  for the
case $M=K$ is characterized as follows:
\begin{thm}\label{thm:DoF-Region}
If the channel matrices $\{\mathbf{H}[n]\}$ is an i.i.d.  process over
time and is also independent and identically distributed across the
receivers, then the $\DoF$ region for the
case $M=K$ is characterized as all positive $K$--tuples $(d_1, d_2,
\ldots, d_K)$ satisfying:
\begin{align}
\sum_{i=1}^K \frac{d_{\pi(i)}}{i}\leq 1
\end{align}
for all permutations $\pi$ of the set $\{1,\ldots, K\}$.
\end{thm}

The achievability result for Theorem \ref{thm:squre} holds for $M
\ge K-j+1$. We have the following achievability result for general
$M,K$ and $j$.
\begin{thm}\label{thm:nonsqure}
Assume that $\mathbf{H}[n]$ is full rank almost surely for each $n$,
and $\{\mathbf{H}[n]\}$ is stationary and ergodic. If
${\DoF}_{j+1}(M,K)$ is achievable for order--$(j+1)$ symbols, then
${\DoF}_{j}(M,K)$  is achievable for order--$j$ symbols, where
\begin{align}\label{eq:itr_nonsquare}
{\DoF}_j(M,K) =  \frac{\frac{ q_j+1 }{j} }{\frac{1}{j} +  \frac{q_j}{j+1 }\frac{1}{{\DoF}_{j+1}(M,K)}},
\end{align}
and $q_j =\min \{ M-1, K-j\}$.
\end{thm}
Starting from  ${\DoF}^*_K(M,K)=1$, which is simply achievable, one can use iterative equation~\eqref{eq:itr_nonsquare} to derive an achievable ${\DoF}_j(M,K)$ with the following closed form
\begin{align}
\label{eq:sub_opt}
& {\DoF}_j(M,K)= \\
\nonumber
&  \frac{\frac{M}{j}}{ \sum_{i=j}^{K-M}
\frac{1}{i} \left( \frac{M-1}{M}\right)^{i-j} + \left(
\frac{M-1}{M}\right)^{K-M-j+1}(\sum_{i=K-M+1}^{K} \frac{1}{i}) },
\end{align}
for the case of $M < K-j+1$. Unlike the case of $M \ge K-j+1$, however,
the expression in (\ref{eq:sub_opt}) does not match the upper bound
in Theorem \ref{thm:outer--bound--Gen}. In particular, this means
that Theorem \ref{thm:nonsqure} does not allow us to characterize
the degrees of freedom $\DoF^*_1(M,K)$ when the number of users $K$ is
greater than the number of transmit antennas $M$. On the other hand,
it is easy to verify that the achievable ${\DoF}_1(M,K)$ in Theorem
\ref{thm:nonsqure} is increasing with $K$, even when $K > M$.
Therefore, unlike the situation with full CSIT, the degrees of
freedom under delayed CSIT is not determined by the minimum of the
number of transmit antennas and the number of receivers.

For the special case of $M=2$ and $K=3$, we obtain an exact
characterization of the degrees of freedom.

\begin{thm}\label{thm:M=2-K=3}
Assume that $\mathbf{H}[n]$ is full rank almost surely for each $n$,
and $\{\mathbf{H}[n]\}$ is stationary and ergodic, then
${\DoF}^*_{1}(2,3)=\frac{3}{2}$.
\end{thm}

\section{Achievable Scheme for Theorem~\ref{thm:squre}}\label{sec:achievable-square}

In this section, we explain the achievable scheme for
Theorem~\ref{thm:squre}. The key is to understand the square case
when $M=K$.  For simplicity, we start with the
cases $M=K=2$ and $M=K=3$.

\subsection{Achievable Scheme for $M=K=2$}

In this subsection, we show that for the case of $M=K=2$,  the ${\DoF}$
of $\frac{4}{3}$ is achievable. We explain the achievable scheme
from three different perspectives:

\begin{enumerate}
\item Exploiting Side-Information
\item Generating Higher-Order Messages
\item Interference Alignment using Outdated CSIT
\end{enumerate}

For notational clarity, in this subsection we will use
$A$ and $B$ to denote the two receivers instead of $1$ and $2$.

\subsubsection{Exploiting Side-Information}

Let $u_r$ and $v_r$ be symbols from two independently encoded Gaussian codewords intended for receiver $r$.  The proposed communication scheme is performed in two phases, which take three time--slots in total:

{\bf Phase One} -- \emph{Feeding the Receivers:} This phase has two time--slots.

The first time slot is dedicated to receiver $A$. The transmitter sends the two symbols, $u_A$ and $v_A$, intended for receiver $A$, i.e.
\begin{align}
\mathbf{x}[1]=
\left[
\begin{array}{c}
u_A\\
v_A
\end{array}
\right].
\end{align}
At the receivers, we have:
\begin{align}
y_A[1]&=h^{\dagger}_{A1}[1]u_A+ h^{\dagger}_{A2}[1] v_A+z_{A}[1],\\
y_B[1]&=h^{\dagger}_{B1}[1]u_A+ h^{\dagger}_{B2}[1] v_A+z_{B}[1].
\end{align}
Both receivers $A$ and $B$ receive  noisy versions of linear combinations of $u_A$ and $u_B$. Receiver $B$ saves the overheard equation for later usage, although it only carries information intended for receiver $A$.

The second time-slot of phase one is dedicated to the second receiver. In this time-slot, the transmitter sends symbols intended for receiver $B$, i.e.
\begin{align}
\mathbf{x}[2]=
\left[
\begin{array}{c}
u_B\\
v_B
\end{array}
\right].
\end{align}
At receivers, we have:
\begin{align}
y_A[2]&=h^{\dagger}_{A1}[2]u_B+ h^{\dagger}_{A2}[2] v_B+z_{A}[2],\\
y_B[2]&=h^{\dagger}_{B1}[2]u_B+ h^{\dagger}_{B2}[2] v_B+z_{B}[2].
\end{align}
Receiver $A$ saves the overheard equation for future usage, although it only carries information intended for receiver $B$.

Let us define short hand notations
\begin{eqnarray*}
L_1(u_A,v_A) & = & h^{\dagger}_{A1}[1]u_A+ h^{\dagger}_{A2}[1] v_A,\\
L_2(u_A,v_A) & = & h^{\dagger}_{B1}[1]u_A+ h^{\dagger}_{B2}[1] v_A,\\
L_3(u_B,v_B) & = & h^{\dagger}_{A1}[2]u_B+ h^{\dagger}_{A2}[2] v_B,\\
L_4(u_B,v_B) & = & h^{\dagger}_{B1}[2]u_B+ h^{\dagger}_{B2}[2] v_B.
\end{eqnarray*}

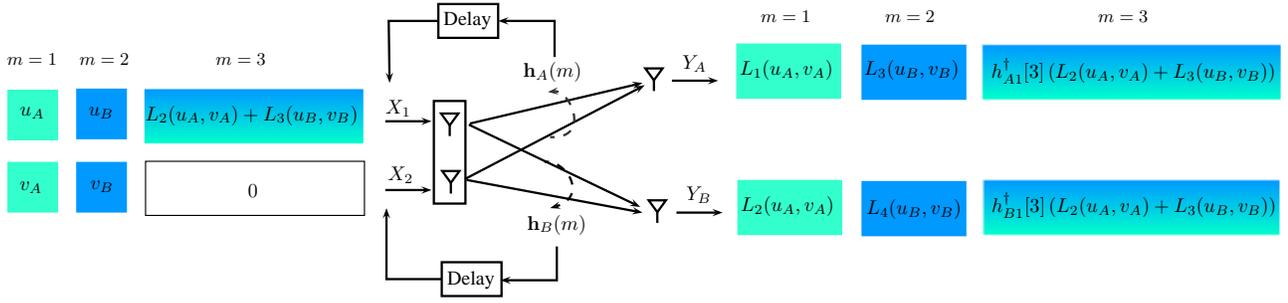
\begin{figure*}[tbhp]
\centering
\scalebox{0.75} 
{
\begin{pspicture}(0,-2.61)(24.344063,2.61)
\definecolor{color251}{rgb}{0.2,0.2,0.2}
\definecolor{color280b}{rgb}{0.2,1.0,0.8}
\definecolor{color302b}{rgb}{0.0,0.6,1.0}
\definecolor{color434f}{rgb}{0.0,1.0,0.8}
\psline[linewidth=0.04cm,arrowsize=0.05291667cm 2.0,arrowlength=1.4,arrowinset=0.4]{->}(6.7621875,0.51)(7.5821877,0.51)
\psline[linewidth=0.04cm,arrowsize=0.05291667cm 2.0,arrowlength=1.4,arrowinset=0.4]{->}(6.7621875,-0.71)(7.5821877,-0.71)
\usefont{T1}{ptm}{m}{n}
\rput(6.9935937,0.74){$X_1$}
\usefont{T1}{ptm}{m}{n}
\rput(7.0335937,-0.44){$X_2$}
\pspolygon[linewidth=0.04](7.7507544,0.6477348)(8.056264,0.6477348)(7.9035096,0.43754685)
\psline[linewidth=0.04cm](7.9035096,0.45756474)(7.9035096,0.24737674)
\pspolygon[linewidth=0.04](7.7707543,-0.37226516)(8.076264,-0.37226516)(7.9235096,-0.5824532)
\psline[linewidth=0.04cm](7.9235096,-0.56243527)(7.9235096,-0.77262324)
\psframe[linewidth=0.04,dimen=outer](8.202188,0.89)(7.6021876,-0.97)
\psline[linewidth=0.04cm,arrowsize=0.05291667cm 2.0,arrowlength=1.4,arrowinset=0.4]{->}(11.842188,1.23)(12.662188,1.23)
\psline[linewidth=0.04cm,arrowsize=0.05291667cm 2.0,arrowlength=1.4,arrowinset=0.4]{->}(11.922188,-1.11)(12.7421875,-1.11)
\usefont{T1}{ptm}{m}{n}
\rput(12.243594,1.54){$Y_A$}
\usefont{T1}{ptm}{m}{n}
\rput(12.323594,-0.8){$Y_B$}
\pspolygon[linewidth=0.04](11.370754,1.4477348)(11.676265,1.4477348)(11.523509,1.2375468)
\psline[linewidth=0.04cm](11.523509,1.2575648)(11.523509,1.0473768)
\pspolygon[linewidth=0.04](11.430755,-0.89226514)(11.736264,-0.89226514)(11.583509,-1.1024531)
\psline[linewidth=0.04cm](11.583509,-1.0824353)(11.583509,-1.2926233)
\psline[linewidth=0.04cm,arrowsize=0.05291667cm 2.0,arrowlength=1.4,arrowinset=0.4]{->}(8.262188,0.47)(11.322187,1.19)
\psline[linewidth=0.04cm,arrowsize=0.05291667cm 2.0,arrowlength=1.4,arrowinset=0.4]{->}(8.182187,-0.51)(11.362187,1.15)
\psline[linewidth=0.04cm,arrowsize=0.05291667cm 2.0,arrowlength=1.4,arrowinset=0.4]{->}(8.2421875,0.45)(11.282187,-1.01)
\psline[linewidth=0.04cm,arrowsize=0.05291667cm 2.0,arrowlength=1.4,arrowinset=0.4]{->}(8.202188,-0.53)(11.322187,-1.11)
\psbezier[linewidth=0.04,linecolor=color251,linestyle=dashed,dash=0.16cm 0.16cm,arrowsize=0.05291667cm 2.0,arrowlength=1.4,arrowinset=0.4]{->}(9.682187,0.23)(10.422188,0.19)(10.042188,1.07)(9.642187,1.03)
\psbezier[linewidth=0.04,linestyle=dashed,dash=0.16cm 0.16cm,arrowsize=0.05291667cm 2.0,arrowlength=1.4,arrowinset=0.4]{<-}(9.682187,-1.03)(10.362187,-0.83)(9.982187,-0.17)(9.582188,-0.21)
\usefont{T1}{ptm}{m}{n}
\rput(9.723594,1.38){$\mathbf{h}_A(m)$}
\usefont{T1}{ptm}{m}{n}
\rput(9.803594,-1.32){$\mathbf{h}_{B}(m)$}
\psframe[linewidth=0.04,dimen=outer](8.782187,2.61)(7.6821876,1.97)
\usefont{T1}{ptm}{m}{n}
\rput(8.217969,2.3){Delay}
\psframe[linewidth=0.04,dimen=outer](8.842188,-1.97)(7.7421875,-2.61)
\usefont{T1}{ptm}{m}{n}
\rput(8.277968,-2.3){Delay}
\psline[linewidth=0.04,arrowsize=0.05291667cm 2.0,arrowlength=1.4,arrowinset=0.4]{->}(7.7621875,-2.29)(6.7821875,-2.29)(6.7821875,-1.03)
\psline[linewidth=0.04,arrowsize=0.05291667cm 2.0,arrowlength=1.4,arrowinset=0.4]{->}(9.762188,1.65)(9.762188,2.29)(8.762188,2.29)
\psline[linewidth=0.04,arrowsize=0.05291667cm 2.0,arrowlength=1.4,arrowinset=0.4]{->}(7.6821876,2.29)(6.8221874,2.29)(6.837275,1.19)
\psline[linewidth=0.04,arrowsize=0.05291667cm 2.0,arrowlength=1.4,arrowinset=0.4]{->}(9.802188,-1.71)(9.802188,-2.31)(8.802188,-2.31)
\psframe[linewidth=0.04,linecolor=color280b,dimen=outer,fillstyle=solid,fillcolor=color280b](0.9621875,1.07)(0.0621875,0.17)
\usefont{T1}{ptm}{m}{n}
\rput(0.50359374,0.66){$u_A$}
\psframe[linewidth=0.04,linecolor=color280b,dimen=outer,fillstyle=solid,fillcolor=color280b](0.9621875,-0.21)(0.0621875,-1.11)
\usefont{T1}{ptm}{m}{n}
\rput(0.48359376,-0.66){$v_A$}
\psframe[linewidth=0.03,linecolor=color280b,dimen=outer,fillstyle=solid,fillcolor=color280b](14.842188,1.89)(12.982187,0.91)
\usefont{T1}{ptm}{m}{n}
\rput(13.873593,1.42){$L_1(u_A, v_A)$}
\psframe[linewidth=0.03,linecolor=color280b,dimen=outer,fillstyle=solid,fillcolor=color280b](14.862187,-0.53)(13.002188,-1.51)
\usefont{T1}{ptm}{m}{n}
\rput(13.893594,-1.0){$L_2(u_A, v_A)$}
\psframe[linewidth=0.04,linecolor=color302b,dimen=outer,fillstyle=solid,fillcolor=color302b](2.1821876,1.09)(1.2821875,0.19)
\usefont{T1}{ptm}{m}{n}
\rput(1.7235937,0.66){$u_B$}
\psframe[linewidth=0.04,linecolor=color302b,dimen=outer,fillstyle=solid,fillcolor=color302b](2.1821876,-0.21)(1.2821875,-1.11)
\usefont{T1}{ptm}{m}{n}
\rput(1.7435937,-0.64){$v_B$}
\psframe[linewidth=0.03,linecolor=color302b,dimen=outer,fillstyle=solid,fillcolor=color302b](17.062187,1.87)(15.202188,0.89)
\usefont{T1}{ptm}{m}{n}
\rput(16.093594,1.42){$L_3(u_B, v_B)$}
\psframe[linewidth=0.03,linecolor=color302b,dimen=outer,fillstyle=solid,fillcolor=color302b](17.062187,-0.55)(15.202188,-1.53)
\usefont{T1}{ptm}{m}{n}
\rput(16.133595,-1.04){$L_4(u_B, v_B)$}
\usefont{T1}{ptm}{m}{n}
\rput(4.413594,-0.72){$0$}
\psframe[linewidth=0.0020,linecolor=white,dimen=outer,fillstyle=gradient,gradlines=2000,gradbegin=color302b,gradend=color434f,gradmidpoint=1.0](6.3821874,1.09)(2.4821875,0.11)
\usefont{T1}{ptm}{m}{n}
\rput(4.393594,0.62){$L_2(u_A,v_A)+ L_3(u_B,v_B)$}
\psframe[linewidth=0.02,dimen=outer](6.4021873,-0.19)(2.5021875,-1.17)
\psframe[linewidth=0.0020,linecolor=white,dimen=outer,fillstyle=gradient,gradlines=2000,gradbegin=color302b,gradend=color434f,gradmidpoint=1.0](22.642187,1.87)(17.342188,0.89)
\usefont{T1}{ptm}{m}{n}
\rput(20.023594,1.4){$h_{A1}^{\dagger} [3]\left(L_2(u_A,v_A)+ L_3(u_B,v_B)\right)$}
\psframe[linewidth=0.0020,linecolor=white,dimen=outer,fillstyle=gradient,gradlines=2000,gradbegin=color302b,gradend=color434f,gradmidpoint=1.0](22.622187,-0.53)(17.362188,-1.51)
\usefont{T1}{ptm}{m}{n}
\rput(20.033594,-0.98){$h_{B1}^{\dagger}[3] \left(L_2(u_A,v_A)+ L_3(u_B,v_B) \right)$}
\usefont{T1}{ptm}{m}{n}
\rput(0.4996875,1.61){\small $m=1$}
\usefont{T1}{ptm}{m}{n}
\rput(1.7796875,1.61){\small $m=2$}
\usefont{T1}{ptm}{m}{n}
\rput(4.1996875,1.61){\small $m=3$}
\usefont{T1}{ptm}{m}{n}
\rput(13.859688,2.37){\small $m=1$}
\usefont{T1}{ptm}{m}{n}
\rput(16.059687,2.37){\small $m=2$}
\usefont{T1}{ptm}{m}{n}
\rput(19.839687,2.37){\small $m=3$}
\end{pspicture}
}

\centering \caption{Achievable Scheme for $M=K=2$}
\label{figure:m2k2}
\end{figure*}

The transmission scheme is summarized in Fig.~\ref{figure:m2k2}. In
this figure, for simplicity, we drop the thermal noise from the
received signals.  We note that, assuming  $\mathbf{H}[1]$ is full
rank, there is a one-to-one map between $(u_A, v_A)$ and
$(L_1(u_A,v_A), L_2(u_A,v_A))$. If receiver $A$ has the  equation
overheard by receiver $B$, i.e. $L_2(u_A, v_A)$, then it has enough
equations to solve for its own symbols $u_A$, and $v_A$. Similarly,
assuming $\mathbf{H}[2]$ is full rank, there is a one-to-one map
between $(u_B, v_B)$ and $(L_3(u_B,v_B), L_4(u_B,v_B))$. If receiver
$B$ has the equation overheard  by receiver $A$, i.e. $L_3(u_B,
v_B)$, then it has enough equations to solve for its own symbols
$u_B$, and $v_B$.

Therefore, the main mission of the second phase is to swap these two overheard equations through the transmitter.


{\bf Phase Two} -- \emph{Swapping Overheard Equations:} This phase  takes
only one time--slot at $n=3$.  At this time, the transmitter sends a
linear combination of the overheard equations, i.e. $L_2(u_A,v_A)$
and $L_3(u_B,v_B)$.  We note that at this time the transmitter is
aware of the CSI at $n=1$ and $n=2$; therefore it can form the
overheard equations $L_2(u_A,v_A)$ and $L_3(u_B,v_B)$.

For example,  $\mathbf{x}[3]$ can be  formed as,
\begin{align}
\mathbf{x}[3]=
\left[
\begin{array}{c}
L_2(u_A,v_A)+L_3(u_B,v_B)\\
0
\end{array}
\right].
\end{align}
At receivers, we have,
\begin{align}
y_A[3]&=h^{\dagger}_{A1}[3] \left( L_2(u_A,v_A)+L_3(u_B,v_B) \right) +z_{A}[3],\\
y_B[3]&=h^{\dagger}_{B1}[3] \left( L_2(u_A,v_A)+L_3(u_B,v_B) \right) +z_{B}[3].
\end{align}

Remember that receiver $A$ already has (a noisy version of)  $L_3(u_B,v_B)$. Thus, together with $y_A[3]$, it can solve for its two symbols $u_A,v_A$.  We have a similar situation for receiver $B$.


\textbf{ Remark:} In this scheme, we assume that in the first time--slot, transmit antenna one sends $u_A$ and transmit antenna two sends $v_A$. However, antenna one and two can send any random linear combination of $u_A$ and $v_A$. Therefore, for example, we can have
\begin{align}
\mathbf{x}[1]= \mathbf{A}[1]
\left[
\begin{array}{c}
u_A\\
v_A
\end{array}
\right],
\end{align}
 where $\mathbf{A}[1] \in \mathbb{C}^{2\times 2 }$ is a randomly selected matrix. Similar statement is true for the second time--slot. At time--slot $n=3$, we send $L_3(u_B,v_B)+L_2(u_A,v_A)$. However, we can send any combination of $L_3(u_B,v_B)$ and $L_2(u_A,v_A)$. In other words,
 \begin{align}
\mathbf{x}[3]= \mathbf{A}[3]
\left[
\begin{array}{c}
L_3(u_B,v_B)\\
L_2(u_A,v_A)
\end{array}
\right],
\end{align}
where  $\mathbf{A}[3] \in \mathbb{C}^{2\times 2 }$ is a randomly
selected matrix. However, we can limit the choice of $\mathbf{A}[3]$
to rank one matrices.

\textbf{Remark:} We note that only the number of independent noisy
equations that each receiver has is important.  As long as the
variance of the noise of each equation is bounded, the $\DoF$ is not
affected.  Therefore, in what follows, we ignore noise and just
focus on the number of independent equations available at each
receiver.

\textbf{Remark:} Note that if the transmitter has $2N$ transmit
antennas, and each of the receivers has $N$ antennas, then we can
follow the same scheme and achieve $\DoF$ of $\frac{4N}{3}$.

\subsubsection{Generating Higher Order Symbols}
We can observe the achievable scheme from another perspective. Remember in the second phase, we send a linear combination of $L_2(u_A,v_A)$ and $L_3(u_B,v_B)$, e.g. $L_2(u_A,v_A)+L_3(u_B,v_B)$, to both receivers. We can consider $L_2(u_A,v_A)+ L_3(u_B,v_B)$ as an\emph{ order--two common symbol}, required by both receivers. Let us define  $u_{AB}=L_2(u_A,v_A)+L_3(u_B,v_B)$. If we have an algorithm which achieves the degrees of freedom of ${\DoF}_2$ for order--two common symbols, then we need $\frac{1}{{\DoF}_2(2,2)}$ time--slots to deliver the common symbol $u_{AB}$ to both receivers. Therefore, in total, we need $2+\frac{1}{{\DoF}_2(2,2)}$ to deliver four symbols $u_A$, $v_A$, $u_B$, and $v_B$ to the designated receivers. Thus, we have,
\begin{align}
{\DoF}_1(2,2)=\frac{4}{2+\frac{1}{{\DoF}_2(2,2)}}.
\end{align}
It is easy to see that we can achieve ${\DoF}_2(2,2)=1$ by simply sending $u_{AB}$ to both receivers in one time--slot. Therefore, ${\DoF}_1(2,2)$ of $\frac{4}{3}$ is achievable.

In summary, phase one takes as input two order--one symbols for each receiver. It takes two time--slots to deliver one desired equation to each of the receivers. Therefore, each receiver needs one more equation to resolve the desired symbols. If the transmitter ignores the overheard equations, we need two more time--slots to deliver one more equation to each receiver and yield the ${\DoF}$ of one. However, by exploiting the overheard equations, we can form a common symbol of order two. Delivering one common symbol of order two to both receivers takes only one time--slot but  it simultaneously provides one useful equation to each of the receivers. Therefore using this scheme, we save one time--slot and achieve ${\DoF}_1(2,2)=\frac{4}{3}$ rather than $\frac{4}{4}$.

\subsubsection{Interference Alignment using Outdated CSIT}
Putting together the symbols received by receiver $A$ over the three time--slots, we have \eqref{eq:super1}. From~\eqref{eq:super1}, it is easy to see that at receiver $A$, the two interference streams $u_B$ and $v_B$ arrived from the same directions $[0, h_{A1}[2], h_{A1}[3]h_{A1}[2]]^{\dagger}$, and therefore $u_B$ and $v_B$ are aligned. Note that the alignment is done using outdated CSIT. By making the interference data symbols aligned at receiver $A$, the two symbols $u_B$ and $v_B$ collapse into one symbol $h^{\dagger}_{A1}[2]u_B+h^{\dagger}_{A2}[2]v_B$.  Eliminating the variable $h^{\dagger}_{A1}[2]u_B+h^{\dagger}_{A2}[2]v_B$ from ~\eqref{eq:super1}, we have~\eqref{eq:super1c}, which is an equation set of the two desired symbols $u_A$ and $v_A$.
It is easy to see that as long as $ h^{\dagger}_{A1}[3] \neq 0 $ and
 $h^{\dagger}_{A1}[1] h^{\dagger}_{B2}[1] - h^{\dagger}_{A2}[1]
 h^{\dagger}_{B1}[1] \neq 0$, then the desired data symbols are not aligned at receiver $A$ and they can be solved for. We note that at $h^{\dagger}_{A1}[1] h^{\dagger}_{B2}[1] - h^{\dagger}_{A2}[1]
 h^{\dagger}_{B1}[1]$ is the determinant of the channel matrix $\mathbf{H}[1]$. Indeed, in this scheme, receiver $A$ borrows the antenna of the second receiver at time--slot $n=1$ to be able to solve for the two symbols.

\begin{figure*}[!t]
\begin{align}\label{eq:super1}
\left[
\begin{array}{c}
y_A[1]\\
y_A[2]\\
y_A[3]
\end{array}
\right]=
\underbrace{
\left[
\begin{array}{cc}
h^{\dagger}_{A1}[1] &  h^{\dagger}_{A2}[1]\\
0 & 0 \\
h^{\dagger}_{A1}[3]  h^{\dagger}_{B1}[1]
 & h^{\dagger}_{A1}[3]  h^{\dagger}_{B2}[1]
\end{array}
\right]}_{\textrm{Rank Two}}
\left[
\begin{array}{c}
u_A\\
v_A\\
\end{array}
\right]
+
\underbrace{
\left[
\begin{array}{cc}
 0 & 0\\
 h^{\dagger}_{A1}[2] &  h^{\dagger}_{A2}[2]\\
h^{\dagger}_{A1}[3]  h^{\dagger}_{A1}[2]
& h^{\dagger}_{A1}[3]  h^{\dagger}_{A2}[2]
\end{array}
\right]}_{\textrm{Rank One}}
\left[
\begin{array}{c}
u_B\\
v_B
\end{array}
\right]
+ \left[
\begin{array}{c}
z_{A}[1]\\
z_{A}[2]\\
z_{A}[3]
\end{array}
\right].
\end{align}

\begin{equation}\label{eq:super1c}
\left[
\begin{array}{c}
y_A[1]\\
y_A[3]-h^{\dagger}_{A1}[3] y_A[2]
\end{array}
\right]=
\left[
\begin{array}{cc}
h^{\dagger}_{A1}[1] &  h^{\dagger}_{A2}[1]\\
h^{\dagger}_{A1}[3] h^{\dagger}_{B1}[1] & h^{\dagger}_{A1}[3] h^{\dagger}_{B2}[2]
\end{array}
\right]
\left[
\begin{array}{c}
u_A\\
v_A\\
\end{array}
\right]+ \left[
\begin{array}{c}
z_{A}[1]\\
z_{A}[3]-h^{\dagger}_{A1}[3] z_{A}[2]
\end{array}
\right].
\end{equation}
\hrulefill
\end{figure*}



\subsection{Achievable Scheme for $M=K=3$} \label{subsec:M=K=3}

In this section, we show how we achieve  ${\DoF}$ of
$\frac{3}{1+\frac{1}{2}+ \frac{1}{3}}=\frac{18}{11}$ for the channel
with a three-antenna transmitter and three single-antenna receivers.
As explained in the previous subsection, we can observe the
achievable scheme from three different perspectives. However, we
find the second perspective simpler to follow. Therefore, in the
rest of the paper, we just explain the algorithm based on the second
perspective.

The achievable scheme has three phases. Phase one takes  order--one
symbols and generates order--two common symbols. Phase two takes
order--two common symbols and generates order--three common symbols.
The last phase takes order three-common symbols and deliver them to
all three receivers.

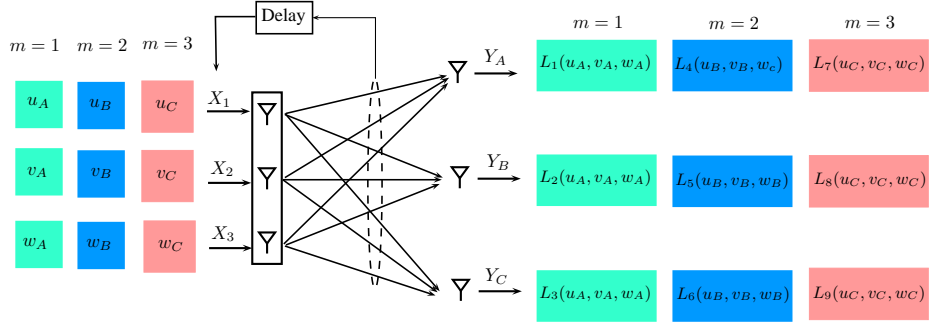
\begin{figure*}[tbhp]
\centering
\scalebox{0.7} 
{
\begin{pspicture}(0,-3.05)(17.97375,3.05)
\definecolor{color9717b}{rgb}{0.2,1.0,0.8}
\definecolor{color9745b}{rgb}{0.0,0.6,1.0}
\definecolor{color9773b}{rgb}{1.0,0.6,0.6}
\psline[linewidth=0.04cm,arrowsize=0.05291667cm 2.0,arrowlength=1.4,arrowinset=0.4]{->}(3.7809374,0.95)(4.6009374,0.95)
\psline[linewidth=0.04cm,arrowsize=0.05291667cm 2.0,arrowlength=1.4,arrowinset=0.4]{->}(3.8009374,-0.41)(4.6209373,-0.41)
\usefont{T1}{ptm}{m}{n}
\rput(4.012344,1.18){$X_1$}
\usefont{T1}{ptm}{m}{n}
\rput(4.072344,-0.14){$X_2$}
\pspolygon[linewidth=0.04](4.7695045,1.0877348)(5.075014,1.0877348)(4.9222593,0.87754685)
\psline[linewidth=0.04cm](4.9222593,0.89756477)(4.9222593,0.68737674)
\pspolygon[linewidth=0.04](4.7695045,-0.13226517)(5.075014,-0.13226517)(4.9222593,-0.34245315)
\psline[linewidth=0.04cm](4.9222593,-0.32243526)(4.9222593,-0.53262323)
\psframe[linewidth=0.04,dimen=outer](5.2209377,1.33)(4.6209373,-1.97)
\psline[linewidth=0.04cm,arrowsize=0.05291667cm 2.0,arrowlength=1.4,arrowinset=0.4]{->}(8.860937,1.67)(9.680938,1.67)
\psline[linewidth=0.04cm,arrowsize=0.05291667cm 2.0,arrowlength=1.4,arrowinset=0.4]{->}(8.920938,-0.33)(9.740937,-0.33)
\usefont{T1}{ptm}{m}{n}
\rput(9.322344,-0.02){$Y_B$}
\pspolygon[linewidth=0.04](8.429504,-0.11226517)(8.735014,-0.11226517)(8.582259,-0.32245317)
\psline[linewidth=0.04cm](8.582259,-0.30243525)(8.582259,-0.51262325)
\usefont{T1}{ptm}{m}{n}
\rput(9.262343,1.98){$Y_A$}
\pspolygon[linewidth=0.04](8.389504,1.8877348)(8.695014,1.8877348)(8.542259,1.6775469)
\psline[linewidth=0.04cm](8.542259,1.6975647)(8.542259,1.4873767)
\psline[linewidth=0.03cm,arrowsize=0.05291667cm 2.0,arrowlength=1.4,arrowinset=0.4]{->}(5.2809377,0.91)(8.340938,1.63)
\psline[linewidth=0.03cm,arrowsize=0.05291667cm 2.0,arrowlength=1.4,arrowinset=0.4]{->}(5.2609377,-0.31)(8.380938,1.59)
\psline[linewidth=0.03cm,arrowsize=0.05291667cm 2.0,arrowlength=1.4,arrowinset=0.4]{->}(5.2609377,0.89)(8.260938,-0.31)
\psline[linewidth=0.03cm,arrowsize=0.05291667cm 2.0,arrowlength=1.4,arrowinset=0.4]{->}(5.2209377,-0.35)(8.240937,-0.35)
\psframe[linewidth=0.03,dimen=outer](5.8009377,3.05)(4.7009373,2.41)
\usefont{T1}{ptm}{m}{n}
\rput(5.2367187,2.74){Delay}
\psline[linewidth=0.02,arrowsize=0.05291667cm 2.0,arrowlength=1.4,arrowinset=0.4]{->}(6.9809375,1.57)(6.9809375,2.73)(5.7809377,2.73)
\pspolygon[linewidth=0.04](4.7695045,-1.3522651)(5.075014,-1.3522651)(4.9222593,-1.5624532)
\psline[linewidth=0.04cm](4.9222593,-1.5424353)(4.9222593,-1.7526232)
\psline[linewidth=0.04cm,arrowsize=0.05291667cm 2.0,arrowlength=1.4,arrowinset=0.4]{->}(3.8209374,-1.65)(4.6409373,-1.65)
\usefont{T1}{ptm}{m}{n}
\rput(4.092344,-1.38){$X_3$}
\psline[linewidth=0.04cm,arrowsize=0.05291667cm 2.0,arrowlength=1.4,arrowinset=0.4]{->}(8.940937,-2.47)(9.760938,-2.47)
\usefont{T1}{ptm}{m}{n}
\rput(9.282344,-2.18){$Y_C$}
\pspolygon[linewidth=0.04](8.449505,-2.2522652)(8.755014,-2.2522652)(8.60226,-2.4624531)
\psline[linewidth=0.04cm](8.60226,-2.4424353)(8.60226,-2.6526232)
\psline[linewidth=0.03cm,arrowsize=0.05291667cm 2.0,arrowlength=1.4,arrowinset=0.4]{->}(5.2609377,0.91)(8.200937,-2.47)
\psline[linewidth=0.03cm,arrowsize=0.05291667cm 2.0,arrowlength=1.4,arrowinset=0.4]{->}(5.2209377,-0.35)(8.160937,-2.49)
\psline[linewidth=0.03cm,arrowsize=0.05291667cm 2.0,arrowlength=1.4,arrowinset=0.4]{->}(5.2409377,-1.57)(8.360937,1.49)
\psline[linewidth=0.03cm,arrowsize=0.05291667cm 2.0,arrowlength=1.4,arrowinset=0.4]{->}(5.2809377,-1.59)(8.200937,-0.43)
\psline[linewidth=0.03cm,arrowsize=0.05291667cm 2.0,arrowlength=1.4,arrowinset=0.4]{->}(5.2609377,-1.61)(8.120937,-2.55)
\psellipse[linewidth=0.03,linestyle=dashed,dash=0.16cm 0.16cm,dimen=outer](6.9709377,-0.42)(0.15,2.01)
\psline[linewidth=0.03,arrowsize=0.05291667cm 2.0,arrowlength=1.4,arrowinset=0.4]{->}(4.7009373,2.75)(3.9409375,2.75)(3.9582102,1.71)
\psframe[linewidth=0.04,linecolor=color9717b,dimen=outer,fillstyle=solid,fillcolor=color9717b](1.0409375,1.53)(0.1409375,0.63)
\usefont{T1}{ptm}{m}{n}
\rput(0.58234376,1.12){$u_A$}
\psframe[linewidth=0.04,linecolor=color9717b,dimen=outer,fillstyle=solid,fillcolor=color9717b](1.0409375,0.25)(0.1409375,-0.65)
\usefont{T1}{ptm}{m}{n}
\rput(0.5623438,-0.2){$v_A$}
\psframe[linewidth=0.04,linecolor=color9717b,dimen=outer,fillstyle=solid,fillcolor=color9717b](1.0409375,-1.13)(0.1409375,-2.03)
\usefont{T1}{ptm}{m}{n}
\rput(0.52234375,-1.56){$w_A$}
\psframe[linewidth=0.03,linecolor=color9717b,dimen=outer,fillstyle=solid,fillcolor=color9717b](12.320937,2.31)(10.040937,1.33)
\usefont{T1}{ptm}{m}{n}
\rput(11.158438,1.87){\small $L_1(u_A, v_A,w_A)$}
\psframe[linewidth=0.03,linecolor=color9717b,dimen=outer,fillstyle=solid,fillcolor=color9717b](12.320937,0.13)(10.040937,-0.85)
\usefont{T1}{ptm}{m}{n}
\rput(11.158438,-0.31){\small $L_2(u_A, v_A,w_A)$}
\psframe[linewidth=0.03,linecolor=color9717b,dimen=outer,fillstyle=solid,fillcolor=color9717b](12.320937,-2.07)(10.040937,-3.05)
\usefont{T1}{ptm}{m}{n}
\rput(11.158438,-2.51){\small $L_3(u_A, v_A,w_A)$}
\psframe[linewidth=0.04,linecolor=color9745b,dimen=outer,fillstyle=solid,fillcolor=color9745b](2.2209375,1.51)(1.3209375,0.61)
\usefont{T1}{ptm}{m}{n}
\rput(1.7623438,1.08){$u_B$}
\psframe[linewidth=0.04,linecolor=color9745b,dimen=outer,fillstyle=solid,fillcolor=color9745b](2.2209375,0.21)(1.3209375,-0.69)
\usefont{T1}{ptm}{m}{n}
\rput(1.7823437,-0.22){$v_B$}
\psframe[linewidth=0.04,linecolor=color9745b,dimen=outer,fillstyle=solid,fillcolor=color9745b](2.2209375,-1.15)(1.3209375,-2.05)
\usefont{T1}{ptm}{m}{n}
\rput(1.7423438,-1.58){$w_B$}
\psframe[linewidth=0.03,linecolor=color9745b,dimen=outer,fillstyle=solid,fillcolor=color9745b](14.900937,2.31)(12.620937,1.33)
\usefont{T1}{ptm}{m}{n}
\rput(13.688437,1.83){\small $L_4(u_B, v_B,w_c)$}
\psframe[linewidth=0.03,linecolor=color9745b,dimen=outer,fillstyle=solid,fillcolor=color9745b](14.900937,0.11)(12.620937,-0.87)
\usefont{T1}{ptm}{m}{n}
\rput(13.758437,-0.37){\small $L_5(u_B, v_B,w_B)$}
\psframe[linewidth=0.03,linecolor=color9745b,dimen=outer,fillstyle=solid,fillcolor=color9745b](14.900937,-2.07)(12.620937,-3.05)
\usefont{T1}{ptm}{m}{n}
\rput(13.778438,-2.53){\small $L_6(u_B, v_B,w_B)$}
\psframe[linewidth=0.0020,linecolor=color9773b,dimen=outer,fillstyle=solid,fillcolor=color9773b](3.5189376,1.528)(2.5429375,0.552)
\usefont{T1}{ptm}{m}{n}
\rput(3.0223436,1.06){$u_{C}$}
\psframe[linewidth=0.0020,linecolor=color9773b,dimen=outer,fillstyle=solid,fillcolor=color9773b](3.5189376,0.208)(2.5429375,-0.768)
\usefont{T1}{ptm}{m}{n}
\rput(3.0223436,-0.26){$v_{C}$}
\psframe[linewidth=0.0020,linecolor=color9773b,dimen=outer,fillstyle=solid,fillcolor=color9773b](3.5589375,-1.132)(2.5829375,-2.108)
\usefont{T1}{ptm}{m}{n}
\rput(3.1023438,-1.6){$w_{C}$}
\psframe[linewidth=0.0020,linecolor=color9773b,dimen=outer,fillstyle=solid,fillcolor=color9773b](17.480938,2.31)(15.220938,1.33)
\usefont{T1}{ptm}{m}{n}
\rput(16.358437,1.85){\small $L_7(u_C, v_C,w_C)$}
\psframe[linewidth=0.0020,linecolor=color9773b,dimen=outer,fillstyle=solid,fillcolor=color9773b](17.480938,0.11)(15.220938,-0.87)
\usefont{T1}{ptm}{m}{n}
\rput(16.378437,-0.35){\small $L_8(u_C, v_C,w_C)$}
\psframe[linewidth=0.0020,linecolor=color9773b,dimen=outer,fillstyle=solid,fillcolor=color9773b](17.480938,-2.03)(15.220938,-3.01)
\usefont{T1}{ptm}{m}{n}
\rput(16.358437,-2.51){\small $L_9(u_C, v_C,w_C)$}
\usefont{T1}{ptm}{m}{n}
\rput(0.51234376,2.22){$m=1$}
\usefont{T1}{ptm}{m}{n}
\rput(1.7923437,2.22){$m=2$}
\usefont{T1}{ptm}{m}{n}
\rput(3.0323439,2.24){$m=3$}
\usefont{T1}{ptm}{m}{n}
\rput(11.212344,2.64){$m=1$}
\usefont{T1}{ptm}{m}{n}
\rput(13.792344,2.62){$m=2$}
\usefont{T1}{ptm}{m}{n}
\rput(16.372343,2.64){$m=3$}
\end{pspicture}
}
\centering \caption{Achievable Scheme for $K=3$: Phase One}
\label{figure:m3k3p1}
\end{figure*}

\textbf{Phase One:} This phase is similar to phase one for the $2$
by $2$ case. It takes three independent symbols for each receiver
and generates three symbols of order two. Assume that $u_r$, $v_r$,
and $w_r$ represent three symbols, independently Gaussian encoded,
for receiver $r$, $r=A,B,C$. Therefore, in total, there are $9$ data
symbols. This phase has three time-slots, where each time--slot is
dedicated to one of the receivers. In the time-slot dedicated to
receiver $A$, the transmitter sends random linear combinations of
$u_A$, $ v_A$, and  $w_A$ over the three antennas. Similarly, in the
time-slot dedicated to receiver $B$, the transmitter sends random
linear combinations of $u_B$, $ v_B$, and  $w_B$ over the three
antennas. In the time-slot dedicated to receiver $C$, the
transmitter sends random linear combinations of $u_C$, $ v_C$, and
$w_C$ over the three antennas. Refer to Fig. \ref{figure:m3k3p1} for
details.

So far the algorithm has taken three time--slots and delivered three desired equations to the designated receivers. Therefore, in terms of counting the desired equations, the algorithm delivers one equation per time--slot which is natural progress for a system without CSIT. If we ignore the overheard equations, then we need six more time--slots to successfully deliver the 9 data streams, which yields the ${\DoF}$ of one. However, as described in the $2$ by $2$ case, the overheard equations can help us to improve the degrees of freedom.

Let us focus on the time-slot dedicated to receiver $A$. Then, we have the following observations:
 \begin{itemize}
 \item The three equations $L_1(u_A,v_A,w_A)$, $L_2(u_A,v_A,w_A)$, and $L_3(u_A,v_A,w_A)$ form three linearly independent equations of $u_A$, $ v_A$, and  $w_A$, almost surely.

 \item If we somehow deliver the overheard equations $L_2(u_A,v_A,w_A)$ and $L_3(u_A,v_A,w_A)$ to receiver $A$, then it has enough equations to solve for $u_A$, $ v_A$, and  $w_A$.

 \item The two overheard equations $L_2(u_A,v_A,w_A)$ and $L_3(u_A,v_A,w_A)$ plus the equation received by receiver $A$ i.e. $L_1(u_A,v_A,w_A)$, fully represent the original data symbols. Therefore, sufficient information to solve for the data symbols is already available at the receivers, but not exactly at the desired receiver.
  \end{itemize}

We have similar observations about the equations received in the time-slots dedicated to receivers $B$ and $C$. Remember that originally the objective was to deliver $u_r$, $v_r$, and  $w_r$ to receiver $r$. After these three transmissions, we can redefine the objective. The new objective is to deliver:

 \begin{itemize}
 \item (i) the overheard equations $L_2(u_A,v_A,w_A)$ and $L_3(u_A,v_A,w_A)$ to receiver $A$,
 \item (ii) the overheard equations $L_4(u_B,v_B,w_B)$ and $L_6(u_B,v_B,w_B)$ to receiver $B$, and
 \item (iii)  the overheard equations $L_7(u_C,v_C,w_C)$ and $L_8(u_C,v_C,w_C)$ to receiver $C$.
\end{itemize}

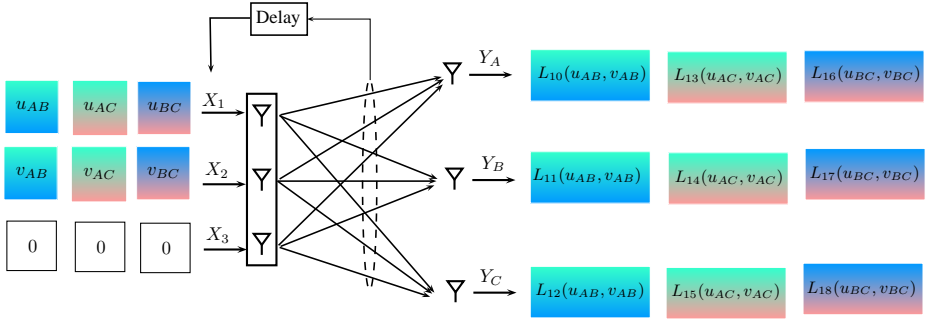
\begin{figure*}[tbhp]
\centering
\scalebox{0.7} 
{
\begin{pspicture}(0,-3.01)(18.61375,3.01)
\definecolor{color292g}{rgb}{0.2,1.0,0.8}
\definecolor{color292f}{rgb}{0.0,0.6,1.0}
\definecolor{color316f}{rgb}{1.0,0.6,0.6}
\psline[linewidth=0.04cm,arrowsize=0.05291667cm 2.0,arrowlength=1.4,arrowinset=0.4]{->}(4.0009375,0.91)(4.8209376,0.91)
\psline[linewidth=0.04cm,arrowsize=0.05291667cm 2.0,arrowlength=1.4,arrowinset=0.4]{->}(4.0209374,-0.45)(4.8409376,-0.45)
\usefont{T1}{ptm}{m}{n}
\rput(4.2323437,1.14){$X_1$}
\usefont{T1}{ptm}{m}{n}
\rput(4.2923436,-0.18){$X_2$}
\pspolygon[linewidth=0.04](4.9895043,1.0477349)(5.2950144,1.0477349)(5.1422596,0.8375468)
\psline[linewidth=0.04cm](5.1422596,0.85756475)(5.1422596,0.6473768)
\pspolygon[linewidth=0.04](4.9895043,-0.17226517)(5.2950144,-0.17226517)(5.1422596,-0.38245317)
\psline[linewidth=0.04cm](5.1422596,-0.36243525)(5.1422596,-0.57262325)
\psframe[linewidth=0.04,dimen=outer](5.4409375,1.29)(4.8409376,-2.01)
\psline[linewidth=0.04cm,arrowsize=0.05291667cm 2.0,arrowlength=1.4,arrowinset=0.4]{->}(9.080937,1.63)(9.900937,1.63)
\psline[linewidth=0.04cm,arrowsize=0.05291667cm 2.0,arrowlength=1.4,arrowinset=0.4]{->}(9.140938,-0.37)(9.9609375,-0.37)
\usefont{T1}{ptm}{m}{n}
\rput(9.542344,-0.06){$Y_B$}
\pspolygon[linewidth=0.04](8.649505,-0.15226518)(8.955014,-0.15226518)(8.802259,-0.36245316)
\psline[linewidth=0.04cm](8.802259,-0.34243527)(8.802259,-0.5526233)
\usefont{T1}{ptm}{m}{n}
\rput(9.482344,1.94){$Y_A$}
\pspolygon[linewidth=0.04](8.609505,1.8477348)(8.915014,1.8477348)(8.7622595,1.6375468)
\psline[linewidth=0.04cm](8.7622595,1.6575648)(8.7622595,1.4473767)
\psline[linewidth=0.03cm,arrowsize=0.05291667cm 2.0,arrowlength=1.4,arrowinset=0.4]{->}(5.5009375,0.87)(8.560938,1.59)
\psline[linewidth=0.03cm,arrowsize=0.05291667cm 2.0,arrowlength=1.4,arrowinset=0.4]{->}(5.4809375,-0.35)(8.600938,1.55)
\psline[linewidth=0.03cm,arrowsize=0.05291667cm 2.0,arrowlength=1.4,arrowinset=0.4]{->}(5.4809375,0.85)(8.480938,-0.35)
\psline[linewidth=0.03cm,arrowsize=0.05291667cm 2.0,arrowlength=1.4,arrowinset=0.4]{->}(5.4409375,-0.39)(8.4609375,-0.39)
\psframe[linewidth=0.03,dimen=outer](6.0209374,3.01)(4.9209375,2.37)
\usefont{T1}{ptm}{m}{n}
\rput(5.456719,2.7){Delay}
\psline[linewidth=0.02,arrowsize=0.05291667cm 2.0,arrowlength=1.4,arrowinset=0.4]{->}(7.2009373,1.53)(7.2009373,2.69)(6.0009375,2.69)
\pspolygon[linewidth=0.04](4.9895043,-1.3922652)(5.2950144,-1.3922652)(5.1422596,-1.6024531)
\psline[linewidth=0.04cm](5.1422596,-1.5824353)(5.1422596,-1.7926233)
\psline[linewidth=0.04cm,arrowsize=0.05291667cm 2.0,arrowlength=1.4,arrowinset=0.4]{->}(4.0409374,-1.69)(4.8609376,-1.69)
\usefont{T1}{ptm}{m}{n}
\rput(4.3123436,-1.42){$X_3$}
\psline[linewidth=0.04cm,arrowsize=0.05291667cm 2.0,arrowlength=1.4,arrowinset=0.4]{->}(9.160937,-2.51)(9.980938,-2.51)
\usefont{T1}{ptm}{m}{n}
\rput(9.502344,-2.22){$Y_C$}
\pspolygon[linewidth=0.04](8.669504,-2.2922652)(8.975015,-2.2922652)(8.82226,-2.502453)
\psline[linewidth=0.04cm](8.82226,-2.4824352)(8.82226,-2.6926231)
\psline[linewidth=0.03cm,arrowsize=0.05291667cm 2.0,arrowlength=1.4,arrowinset=0.4]{->}(5.4809375,0.87)(8.420938,-2.51)
\psline[linewidth=0.03cm,arrowsize=0.05291667cm 2.0,arrowlength=1.4,arrowinset=0.4]{->}(5.4409375,-0.39)(8.380938,-2.53)
\psline[linewidth=0.03cm,arrowsize=0.05291667cm 2.0,arrowlength=1.4,arrowinset=0.4]{->}(5.4609375,-1.61)(8.580937,1.45)
\psline[linewidth=0.03cm,arrowsize=0.05291667cm 2.0,arrowlength=1.4,arrowinset=0.4]{->}(5.5009375,-1.63)(8.420938,-0.47)
\psline[linewidth=0.03cm,arrowsize=0.05291667cm 2.0,arrowlength=1.4,arrowinset=0.4]{->}(5.4809375,-1.65)(8.340938,-2.59)
\psellipse[linewidth=0.03,linestyle=dashed,dash=0.16cm 0.16cm,dimen=outer](7.1909375,-0.46)(0.15,2.01)
\psline[linewidth=0.03,arrowsize=0.05291667cm 2.0,arrowlength=1.4,arrowinset=0.4]{->}(4.9209375,2.71)(4.1609373,2.71)(4.1782103,1.67)
\psframe[linewidth=0.0020,linecolor=white,dimen=outer,fillstyle=gradient,gradlines=2000,gradbegin=color292g,gradend=color292f,gradmidpoint=1.0](1.2779375,1.527)(0.2639375,0.513)
\usefont{T1}{ptm}{m}{n}
\rput(0.8023437,1.04){$u_{AB}$}
\psframe[linewidth=0.02,dimen=outer](1.2509375,-1.14)(0.2909375,-2.1)
\usefont{T1}{ptm}{m}{n}
\rput(0.79234374,-1.62){$0$}
\psframe[linewidth=0.0020,linecolor=white,dimen=outer,fillstyle=gradient,gradlines=2000,gradbegin=color292g,gradend=color292f,gradmidpoint=1.0](1.2579376,0.267)(0.2439375,-0.747)
\usefont{T1}{ptm}{m}{n}
\rput(0.78234375,-0.22){$v_{AB}$}
\psframe[linewidth=0.0020,linecolor=white,dimen=outer,fillstyle=gradient,gradlines=2000,gradbegin=color292g,gradend=color292f,gradmidpoint=1.0](12.520938,2.11)(10.240937,1.13)
\usefont{T1}{ptm}{m}{n}
\rput(11.418438,1.65){\small $L_{10}(u_{AB},v_{AB})$}
\psframe[linewidth=0.0020,linecolor=white,dimen=outer,fillstyle=gradient,gradlines=2000,gradbegin=color292g,gradend=color292f,gradmidpoint=1.0](12.520938,0.17)(10.240937,-0.81)
\usefont{T1}{ptm}{m}{n}
\rput(11.318438,-0.29){\small $L_{11}(u_{AB},v_{AB})$}
\psframe[linewidth=0.0020,linecolor=white,dimen=outer,fillstyle=gradient,gradlines=2000,gradbegin=color292g,gradend=color292f,gradmidpoint=1.0](12.520938,-2.01)(10.240937,-2.99)
\usefont{T1}{ptm}{m}{n}
\rput(11.378437,-2.47){\small $L_{12}(u_{AB},v_{AB})$}
\psframe[linewidth=0.0020,linecolor=white,dimen=outer,fillstyle=gradient,gradlines=2000,gradbegin=color292g,gradend=color316f,gradmidpoint=1.0](2.5969374,1.526)(1.5449375,0.474)
\usefont{T1}{ptm}{m}{n}
\rput(2.0823438,1.02){$u_{AC}$}
\psframe[linewidth=0.02,dimen=outer](2.5509374,-1.16)(1.5909375,-2.12)
\usefont{T1}{ptm}{m}{n}
\rput(2.0923438,-1.64){$0$}
\psframe[linewidth=0.0020,linecolor=white,dimen=outer,fillstyle=gradient,gradlines=2000,gradbegin=color292g,gradend=color316f,gradmidpoint=1.0](2.5769374,0.266)(1.5249375,-0.786)
\usefont{T1}{ptm}{m}{n}
\rput(2.0623438,-0.24){$v_{AC}$}
\psframe[linewidth=0.0020,linecolor=white,dimen=outer,fillstyle=gradient,gradlines=2000,gradbegin=color292g,gradend=color316f,gradmidpoint=1.0](15.120937,2.07)(12.840938,1.09)
\psframe[linewidth=0.0020,linecolor=white,dimen=outer,fillstyle=gradient,gradlines=2000,gradbegin=color292g,gradend=color316f,gradmidpoint=1.0](15.140938,0.15)(12.860937,-0.83)
\psframe[linewidth=0.0020,linecolor=white,dimen=outer,fillstyle=gradient,gradlines=2000,gradbegin=color292g,gradend=color316f,gradmidpoint=1.0](15.100938,-2.03)(12.820937,-3.01)
\usefont{T1}{ptm}{m}{n}
\rput(13.998438,1.61){\small $L_{13}(u_{AC},v_{AC})$}
\usefont{T1}{ptm}{m}{n}
\rput(14.018437,-0.33){\small $L_{14}(u_{AC},v_{AC})$}
\usefont{T1}{ptm}{m}{n}
\rput(13.958438,-2.49){\small $L_{15}(u_{AC},v_{AC})$}
\psframe[linewidth=0.0020,linecolor=white,dimen=outer,fillstyle=gradient,gradlines=2000,gradbegin=color292f,gradend=color316f,gradmidpoint=1.0](3.7979374,1.507)(2.7839375,0.493)
\usefont{T1}{ptm}{m}{n}
\rput(3.3023438,1.02){$u_{BC}$}
\psframe[linewidth=0.02,dimen=outer](3.7909374,-1.18)(2.8309374,-2.14)
\usefont{T1}{ptm}{m}{n}
\rput(3.3323438,-1.66){$0$}
\psframe[linewidth=0.0020,linecolor=white,dimen=outer,fillstyle=gradient,gradlines=2000,gradbegin=color292f,gradend=color316f,gradmidpoint=1.0](3.7779374,0.267)(2.7639375,-0.747)
\usefont{T1}{ptm}{m}{n}
\rput(3.2823439,-0.22){$v_{BC}$}
\psframe[linewidth=0.0020,linecolor=white,dimen=outer,fillstyle=gradient,gradlines=2000,gradbegin=color292f,gradend=color316f,gradmidpoint=1.0](17.720938,2.09)(15.440937,1.11)
\psframe[linewidth=0.0020,linecolor=white,dimen=outer,fillstyle=gradient,gradlines=2000,gradbegin=color292f,gradend=color316f,gradmidpoint=1.0](17.740938,0.23)(15.4609375,-0.75)
\psframe[linewidth=0.0020,linecolor=white,dimen=outer,fillstyle=gradient,gradlines=2000,gradbegin=color292f,gradend=color316f,gradmidpoint=1.0](17.700937,-1.95)(15.420938,-2.93)
\usefont{T1}{ptm}{m}{n}
\rput(16.618437,1.65){\small $L_{16}(u_{BC},v_{BC})$}
\usefont{T1}{ptm}{m}{n}
\rput(16.598438,-0.25){\small $L_{17}(u_{BC},v_{BC})$}
\usefont{T1}{ptm}{m}{n}
\rput(16.538437,-2.43){\small $L_{18}(u_{BC},v_{BC})$}
\end{pspicture}
}
\centering
\caption{Achievable Scheme for $K=3$: Phase Two}
\label{figure:m3k3p2}
\end{figure*}

Let us define $u_{AB}$ as a random linear combination of $L_2(u_A,v_A,w_A)$ and $L_4(u_B,v_B,w_B)$. To be specific,  let $u_{AB}= L_2(u_A,v_A,w_A)+L_4(u_B,v_B,w_B)$. Then we have the following observations:
\begin{itemize}
\item If receiver $A$ has $u_{AB}$, then it can use the saved overheard equation $L_4(u_B,v_B,w_B)$ to obtain $L_2(u_A,v_A,w_A)$. Remember $L_2(u_A,v_A,w_A)$ is a desired equation for receiver $A$.
\item If receiver $B$ has $u_{AB}$, then it can used the saved overheard equation $L_2(u_A,v_A,w_A)$ to obtain $L_4(u_B,v_B,w_B)$. Remember $L_4(u_B,v_B,w_B)$ is a desired equation for receiver $B$.
\end{itemize}
Therefore, $u_{AB}$  is desired by both receivers $A$ and $B$. Similarly, we define $u_{AC}= L_3(u_A,v_A,w_A)+L_7(u_C,v_C,w_C)$, which is desired by receivers $A$ and $C$, and  define  $u_{BC}= L_6(u_B,v_B,w_B)+L_8(u_C,v_C,w_C)$, which is desired by receivers $B$ and $C$. We note that if receiver $A$ has $u_{AB}$ and $u_{AC}$, then it has enough equations to solve the original data symbols $u_A$, $ v_A$, and  $w_A$. Similarly, it is enough that receiver $B$ has $u_{AB}$ and $u_{BC}$, and  receiver $C$ has $u_{AC}$ and $u_{BC}$. Therefore, again, we can redefine the objective as delivering $u_{AB}$ to receivers  $A$ and $B$, $u_{AC}$ to receivers $A$ and $C$, and $u_{BC}$ to receivers $B$ and $C$. Suppose now we have an algorithm that can achieve ${\DoF}_2(3,3)$ degrees of freedom for order--two common symbols. Then, the total time to deliver the original $9$ data symbols is the initial three time--slots of sending linear combinations of the $9$ symbols plus $\frac{3}{{\DoF}_2(3,3)}$  time--slots to deliver the three order--two symbols generated. Therefore, the overall DoF to send the order--1 symbols is given by
\begin{align}\label{DoF_1(3,3)}
{\DoF}_1(3,3)=\frac{9}{3+ \frac{3}{{\DoF}_2(3,3)}}.
\end{align}

It is trivially easy to achieve ${\DoF}_2(3,3)=1$,   which yields ${\DoF}_1(3,3)$ of $\frac{3}{2}$. However, as we will elaborate in the following, we can do better.

%
%

\textbf{Phase Two:} Phase one of the algorithm takes order--one symbols and generates order--two symbols to be delivered. Phase two takes order--two symbols, and generates order--three symbols. Phases two and three together can also be viewed as an algorithm which delivers order-two common symbols.

Assume that $u_{AB}$ and $v_{AB}$ represent two symbols that are desired by both receivers $A$ and $B$. Similarly, $u_{AC}$ and $v_{AC}$ are required by both receivers $A$ and $C$, and $u_{BC}$ and $v_{BC}$ are required by both receivers $B$ and $C$. Therefore, in total, there are 6 order--two symbols. We notice that phase one generates only three order--two symbols. To provide 6 order--two symbols, we can simply repeat phase one twice with new input symbols. Phase two takes three time-slots, where each time--slot is dedicated to one pair of the receivers. In the time-slot dedicated to receivers $A$ and $B$, the transmitter sends random linear combinations $u_{AB}$ and $v_{AB}$ from two of the transmit antennas.  We have analogous transmissions  in the other two time--slots. For details, see Fig.~\ref{figure:m3k3p2}.

In Fig.~~\ref{figure:m3k3p2}, we focus on  the first time--slot dedicated to both users $A$ and $B$. Then, we have the following important observations:
\begin{itemize}
\item $L_{10}(u_{AB}, v_{AB})$ and  $L_{12}(u_{AB}, v_{AB})$ form two linearly independent equations of  $u_{AB}$ and $v_{AB}$, almost surely.
\item Similarly, $L_{11}(u_{AB}, v_{AB})$ and $L_{12}(u_{AB}, v_{AB})$ form two linearly independent equations of  $u_{AB}$ and $v_{AB}$, almost surely.
\item  If $L_{12}(u_{AB}, v_{AB})$ is somehow delivered to both receivers $A$ and $B$, then both receivers have enough equations to solve for $u_{AB}$ and $v_{AB}$. Therefore, $L_{12}(u_{AB}, v_{AB})$, which is overheard and saved by receiver $C$, is {\em simultaneously} useful for receivers $A$ and $B$.
\end{itemize}
We have similar observations about the received equations in the other two time-slots. Therefore, after these three time-slots, we can redefine the objective of the rest of the algorithm as delivering
\begin{itemize}
\item (i) $L_{12}(u_{AB}, v_{AB})$ to receivers $A$ and $B$,
\item (ii) $L_{14}(u_{AC}, v_{AC})$ to receivers $A$ and $C$, and
\item (iii) $L_{16}(u_{BC}, v_{BC})$ to receivers $B$ and $C$.
\end{itemize}

Let us define $u_{ABC}$ and $v_{ABC}$ as any two linearly independent combinations of $L_{12}(u_{AB}, v_{AB})$ and $L_{14}(u_{AC}, v_{AC})$, and $L_{16}(u_{BC}, v_{BC})$:
\begin{align*}
 u_{ABC}  =    \alpha_1 L_{12}(u_{AB}, v_{AB}) & + \alpha_2 L_{14}(u_{AC}, v_{AC}) \\
 &
 + \alpha_3 L_{16}(u_{BC}, v_{BC}), \\
 v_{ABC}  =   \beta_1 L_{12}(u_{AB},v_{AB}) & +   \beta_2 L_{14}(u_{AC}, v_{AC}) \\
& + \beta_3 L_{16}(u_{BC}, v_{BC}),
\end{align*}
where the constants $\alpha_i$ and $\beta_i$, $i=1,2,3$, have been shared with receivers.  If we somehow deliver $u_{ABC}$ and $v_{ABC}$ to receiver $A$, then together with its saved overheard equation $L_{16}(u_{BC}, v_{BC})$, receiver $A$ has 3 linearly independent equations to solve for $L_{12}(u_{AB}, v_{AB})$ and $L_{14}(u_{AC}, v_{AC})$. Then, it has enough equations to solve for $u_{AB}$, $v_{AB}$, $u_{AC}$, and $v_{AC}$.
We have the similar situation for receivers $B$  and $C$. Therefore, it is enough to deliver $u_{ABC}$ and $v_{ABC}$ to all three receivers. If we have an algorithm that can provide ${\DoF}_3(3,3)$ degrees of freedom to deliver order-three common symbols, then the total time to deliver the original $6$ order--two common symbols is $3+ \frac{2}{{\DoF}_3(3,3)}$, taking into account the first three transmissions (described in Fig. \ref{figure:m3k3p2}). Therefore, we have
\begin{align}\label{DoF_2(3,3)}
{\DoF}_2(3,3)=\frac{6}{3+ \frac{2}{{\DoF}_3(3,3)}}.
\end{align}

%
%

\textbf{Phase Three:} Phase Three transmits order--three common symbols. This phase is very simple. Assume that $u_{ABC}$ is required by all three receivers. Then, the transmitter can use only one transmit antenna and send $u_{ABC}$. All three receivers will receive a noisy version of $u_{ABC}$. Therefore, we use one time--slot to send one order--three symbol. Therefore, ${\DoF}_3(3,3)=1$.
Then, from \eqref{DoF_1(3,3)} and \eqref{DoF_2(3,3)}, we conclude that   ${\DoF}_1(3,3)=\frac{18}{11}$ and ${\DoF}_2(3,3)=\frac{6}{5}$.

\subsection{General Proof of Achievability for Theorem~\ref{thm:squre}}

In this section, we explain the achievable scheme for the general
case in Theorem \ref{thm:squre}.

First we focus on the general $M=K$ square case.
The algorithm is based on a
concatenation of $K$ phases. Phase $j$ takes symbols of order $j$
and generates symbols of  order $j+1$. For $j=K$, the phase is
simple and generates no more symbols. For each $j$, we can also
view phases $j,j+1, \ldots K$ together, as an algorithm whose job
is to deliver common symbols of order $j$ to the receivers.

The $j^{\textrm{th}}$ phase takes $(K-j+1) {K \choose j}$ common symbols of order $j$, and yields $j {K \choose {j+1}}$  symbols of order $j+1$. This phase has ${K \choose j}$ time-slots, with each time-slot dedicated to a subset $\mathcal{S}$ of receivers,   $|\mathcal{S}|=j$. We denote the time-slot dedicated to the subset $\mathcal{S}$ by $t_\mathcal{S}$. In this time-slot, the transmitter sends random linear combinations of the $K-j+1$ symbols $u_{\mathcal{S},1}, u_{\mathcal{S},2}, \ldots, u_{\mathcal{S},K-j+1}$, desired by all the receivers in $\mathcal{S}$. The transmitter utilizes $K-j+1$ of the transmit antennas.

The linear combination of the transmitted symbols received by receiver $r$ is denoted by $L_{\mathcal{\mathcal{S}},r}$. Let us focus on the linear combinations of the transmitted symbols  received by all receivers, in time--slot $t_\mathcal{S}$. We have the following observations:
\begin{itemize}
\item For every $r \in \mathcal{S}$, the $K-j+1$ equations consisting of  one equation $L_{\mathcal{\mathcal{S}},r}$ and the $K-j$ overheard equations:  $\{L_{\mathcal{\mathcal{S}},r'}: r' \in \mathcal{E} \backslash \mathcal{S}\}$ are linearly independent equations of the $K-j+1$ symbols $u_{\mathcal{S},1}, u_{\mathcal{S},2}, \ldots, u_{\mathcal{S},K-j+1}$. This relies on the fact that the transmitter uses  $K-j+1$ transmit antennas.

\item For any $r$, $r \in \mathcal{S}$, if we somehow deliver the $K-j$ equations  $\{L_{\mathcal{\mathcal{S}},r'}: r' \in \mathcal{E} \backslash \mathcal{S}\}$ to receiver $r$, then receiver $r$ has  $K-j+1$ linearly independent equations to solve for all $K-j+1$ symbols $u_{\mathcal{S},1}, u_{\mathcal{S},2}, \ldots, u_{\mathcal{S},K-j+1}$.

\item Having the above two observations, we can say that the overheard equation by receiver $r'$,  $r' \in \mathcal{E} \backslash \mathcal{S}$,  is {\em simultaneously} useful for all receivers in $\mathcal{S}$.
\end{itemize}

After repeating the above transmission for all $\mathcal{S}$, where $\mathcal{S}\subset \mathcal{E}$ and $|\mathcal{S}|=j$, then we have another important observation. Consider any subset $\T$ of  receivers, where $|\T|=j+1$. Then each receiver $r$, $r \in \T$, has an overheard equation $L_{\T \backslash \{ r \}, r}$, which is simultaneously useful for all the receivers in $\T \backslash \{ r \}$. We note that the transmitter is aware of these overheard equations. For every $\T \subset \mathcal{E}$, $|\T|=j+1$, the transmitter forms $j$ random linear combinations of $L_{\T \backslash \{ r \}, r}$, $r \in  \T$, denoted by $u_{\T,1},  u_{\T,2}, \ldots, u_{\T,j}$. We note that $u_{\T,\xi}$, $1 \leq  \xi \leq j$, is simultaneously useful for all receivers in $\T$. Indeed, each receiver $r$ in $\T$ can subtract the contribution of $L_{ \T \backslash \{ r \}, r }$ from $u_{\T,\xi}$, $\xi = 1, \ldots , j$, and form $j$ linearly independent combinations of $L_{\T \backslash \{ r \},r}$, $r \in  \T \backslash \{ r \}$. Using the above procedure, the transmitter generates $j {K \choose {j+1}}$ symbols of order $j+1$. The important observation is that if these $j {K \choose {j+1}}$ symbols are delivered to the designated receivers, then each receiver will have enough equations to solve for all of the original common symbols of order $j$. Delivering $j {K \choose {j+1}}$ order--$(j+1)$ symbols takes $\frac{j {K \choose {j+1}}}{{\DoF}_{j+1}(K,K)}$ using an algorithm that provides ${\DoF}_{j+1}(K,K)$ degrees of freedom for order--$(j+1)$ symbols. Since the phase starts with $(K-j+1) {K \choose j}$ symbols of order $j$, and takes ${K \choose j}$ time--slots, and generates $j {K \choose {j+1}}$ symbols with order $j+1$, we have
\begin{align}
{\DoF}_j(K,K)=\frac{   (K-j+1) {K \choose j} }{{K \choose j} +  \frac{j {K \choose {j+1}}}{{\DoF}_{j+1}(K,K)} },
\end{align}
or
\begin{align}
\frac{ K-j+1 }{j} \frac{1}{{\DoF}_j(K,K)}=\frac{1}{j} +  \frac{K-j}{j+1 }\frac{1}{{\DoF}_{j+1}(K,K)} .
\end{align}


It is also easy to see that ${\DoF}_K(K)=1$ is achievable. Solving the recursive equation, we have
\begin{align}
\DoF_j(K,K)= \frac{K-j+1}{j} \frac{1}{
\frac{1}{j}+\frac{1}{j+1}+\ldots+\frac{1}{K} }.
\end{align}
In particular,
\begin{align}
{\DoF}_1(K,K)= \frac{K}{1+\frac{1}{2}+\ldots+\frac{1}{K}}.
\end{align}
Therefore the achievablity of Theorem~\ref{thm:squre} in the square
case has been established.

Now observe that in the above algorithm, phase $j$ only requires the
use of $K-j+1$ transmit antennas, not all $K$ of the transmit
antennas. Moreover, common symbols of order $j$ are delivered using
phases $j,j+1,\ldots, K.$  Hence, we conclude that the degree of
freedom of order--$j$ messages achieved above in the square system
can actually be achieved in a system with less transmit antennas as
long as $M \ge K-j+1$.  This proves Theorem \ref{thm:squre} in the
rectangular case as well.

{\bf Remark:} We note that if the transmitter has $KM$ transmit
antennas, and each of the $K$ receivers has $N$ receive antennas,
then the ${\DoF}_1$ of
$$\frac{KN}{1+\frac{1}{2}+\ldots+\frac{1}{K}}$$ is achievable. More
generally, in this channel, for order--$j$ symbols, the ${\DoF}_j$
of
$$\frac{K-j+1}{j}\frac{N}{\frac{1}{j}+\ldots+\frac{1}{K}}$$ is achievable.

\subsection{Implementation Issues}

For simplicity, the proposed scheme has been presented in a symbol--by--symbol based format. However, this scheme can be implemented in a block--by--block fashion as well. This would allow us to exploit the coherence of the channel over time and frequency to reduce channel training and feedback overhead.
To be specific, let us again focus on the case of $M=K=2$.

Consider a block of time-frequency resources,  consecutive  in time and frequency. Let us assume that in the first phase of the scheme, we dedicate half of these resources to receiver $A$ and the other half to receiver $B$. To start the second phase, the transmitter needs to know  channel coefficients during the first phase.  For example, if the lengths of the block in time and frequency are respectively less than coherent time and  bandwidth of the channel, then during the first phase the channel coefficients are  (almost) constant. Therefore, to start the second phase, the transmitter needs only to know the four channel coefficients. Let us denote the coherent time  and bandwidth by $T_c$ and $W_c$ respectively. Then, for each $T_cW_c$ time-frequency resources, the transmitter needs to dedicate at least two time-frequency resources to send orthogonal pilot signals and learn four coefficients through feedback. Then, the transmitter uses the remaining resources to send $2T_cW_c-2$ order--one symbols. Remember that the transmitter is also required to report the channel coefficients of each receiver to the other receiver. Since each receiver knows its own channel state information,  the transmitter can exploit that and send to both receivers the two symbols of $h_{A1}[1]+h_{B1}[1]$ and $h_{A2}[1]+h_{B2}[1]$, as the symbols of order--two in the second phase.  Therefore, the second phase takes $\frac{2T_cW_c-2}{4}+2$ resource units for order--two messages. Following the above argument, the scheme can achieve ${\DoF}$  of $\frac{2T_cW_c-2}{\frac{3}{2}T_cW_c+0.5}$. If $ T_c W_c \gg 1$, as in most wireless channels, then the degree of freedom is close to $4/3$.

\section{Outer-Bound} \label{sec:outer} 
In this section, we aim to prove
Theorem~\ref{thm:outer--bound--Gen}. In this theorem, we focus on
the degrees of freedom of the channel for order--$j$ messages.
Therefore, we assume for every subset $\mathcal{S}$
with cardinality $j$ of receivers, the transmitter has a message
$W_{\mathcal{S}}$, with rate $R_{\mathcal{S}}$ and degrees of
freedom $d_{\mathcal{S}}$.

Remember in Section~\ref{sec:problem_formulation}, we assume that
the channel state information is available to all nodes with one
time-unit delay. As an outer-bound, we consider the capacity of a
channel in which the channel state information at time $n$   is
available to all {\em receivers} instantaneously at time $n$.
Therefore, at time $n$,  receiver $r$ has  $\left(y_r[t],
\mathbf{H}[t] \right)$, $t=1,\ldots,m$, for any $r$, $1 \leq r \leq
K$.  On the other hand,  the transmitter has not only the channel
state information, but also received signals, both with one unit
delay. Therefore, at time $n$, the transmitter has
$\left(y_1[t],\ldots, y_K[t], \mathbf{H}[t]\right)$,
$t=1,\ldots,m-1$. Now, we improve the resultant channel even further
as follows.

Consider a permutation $\pi$ of the set $\mathcal{E} = \{1, 2,
\ldots, K\}$.  We form a $K$--receiver broadcast channel, by giving
the output of the receiver $\pi(i)$ to the receivers $\pi(j)$,
$j=i+1,\ldots, K$, for all $i=1,\ldots,K-1$. Therefore, we have an
upgraded broadcast channel, referred to as \emph{improved channel} with
$K$ receivers as   $\left(y_{\pi(1)}[n] , \mathbf{H}[n] \right)$,
$\left(y_{\pi(1)}[n] , y_{\pi(2)}[n], \mathbf{H}[n] \right)$,
$\ldots$, $\left(y_{\pi(1)}[n] , y_{\pi(2)}[n], \ldots,
y_{\pi(K)}[n],\mathbf{H}[n] \right)$. We denote the capacity of the
resultant channel as  $\mathcal{C}_{\textrm{Improved}} (\pi)$.
Denoting the capacity of the original channel with $\mathcal{C}$,
we  obviously have $\mathcal{C} \subset
\mathcal{C}_{\textrm{Improved}}(\pi)$. Moreover,  it is easy to see
that the improved channel is physically degraded.

In the improved channel, consider message $W_{\mathcal{S}}$, which
is required by all $j$ receivers listed in $\mathcal{S}$. Let $i^*$
be the smallest integer where $\pi(i^*) \in \mathcal{S}$. Then, due
to the degradedness of the channel, if $W_{\mathcal{S}}$ is decoded
by receiver $\pi(i^*)$,  then it can be decoded by all other
receivers in $\mathcal{S}$. Therefore, we can assume that
$W_{\mathcal{S}}$ is just required by receiver $\pi(i^*)$.  Using
this argument, we can simplify the messages requirements from
order--$j$ common messages to pure private messages as follows:
receiver $\pi(1)$ requires all messages $W_{\mathcal{S}}$, where
$\pi(1)\in \mathcal{S}$ and $\mathcal{S} \in \mathcal{E}$.
Similarly, receiver $\pi(2)$ requires all messages
$W_{\mathcal{S}}$, where $\pi(2)\in \mathcal{S}$ and $\mathcal{S}
\subset \mathcal{E} \backslash \{ \pi(1)\}$. We follow the same
argument for all receivers.

According to~\cite{Elgamal_BC_Feedback}, feedback does not improve the capacity of the physically degraded broadcast channels. Consequently, we focus on the capacity region of the improved channel without feedback, and with the new private message set. On the other hand, for broadcast channels without feedback, the capacity region is only a function of marginal distributions. Therefore, we can ignore the coupling between the receivers in the improved channel. Thus, we have a broadcast channel where receiver $\pi(i)$ has $i$ antennas, and the distributions of the channels between the transmitter and any of the receive antennas are identical. Moreover receiver $\pi(i)$ is interested in all messages $W_{\mathcal{S}}$, where  $\pi(i)\in \mathcal{S}$, $| \mathcal{S}|=j$, and  $\mathcal{S} \subset \mathcal{E} \backslash \{ \pi(1), \pi(2), \ldots, \pi(i-1)\}$.

 Therefore, according to~\cite{NOCSIT_Jafar_J}, extended by~\cite{vaze_noCSIT}, one can conclude that

\begin{align} \label{ineq:outer--bound}
\sum_{i=1}^{K-j+1}  \frac{1}{\min  \{ i, M \} }
\displaystyle{ \sum_{
\substack{
\mathcal{|S|}=j \\
\mathcal{S} \subset \mathcal{E}\backslash \{ \pi(1), \ldots \pi(i-1) \} \\ \pi(i) \in \mathcal{S}
}}
d_{\mathcal{S}} }\leq 1.
\end{align}



By applying the same procedure for any permutation of the set $\{1, 2, \ldots, K\}$ and then adding all of the $K!$ resulting inequalities, the theorem follows.

\section{The $\DoF$ region for $K=M$}\label{Sec:DoF-Region}
In this section, we prove Theorem \ref{thm:DoF-Region} which characterizes the $\DoF$ region of the channel for the case $M=K$.

We note that the region of Theorem~\ref{thm:DoF-Region} is the
polyhedron proposed by the outer--bound \eqref{ineq:outer--bound} for order--one messages where $M=K$.
Here, we show by induction on $K$ that the region is achievable. The
hypothesis is clearly true for $K=1$. Now assume that the hypothesis is
true for $K =1, \ldots , k-1$. Consider the case when $K=k$. First
we argue that any point $(\td_1, \td_2, \ldots, \td_k)$ in the
polyhedron such that $\td_i > 0$ for all $i$ and $\td_i \not =
\td_j$ for some $i,j$ cannot be a corner point of the polyhedron.
Without loss of generality, we can assume that the coordinates of
such a point is ordered in a non-decreasing order, since the
polyhedron is invariant to permutation of coordinates. Let $i_1,
i_2$ be such that either $0< \td_1 = \ldots = \td_{i_1} <
\td_{i_1+1} = \ldots = \td_{i_2} < \td_{i_2 +1}$, or $0 < \td_1 =
\ldots = \td_{i_1} < \td_{i_1+1} = \ldots = \td_{i_2}$ and $i_2 =k$.
Now a direct calculation shows that $\pi$ is a permutation of $\{1,
\ldots, k\}$ which maximizes:
$$ \sum_{i=1}^k \frac{\td_{\pi(i)}}{i}$$
among all permutations if and only if $\td_{\pi(i)} > \td_{\pi(j)}$
whenever $i < j$. This means that the only constraints, if any, of
the polyhedron that $(\td_1, \ldots, \td_k)$ satisfies with equality
correspond to permutations satisfying $\pi(j) \in \{1, \ldots
i_1\}$ for all $j \in \{k-i_1+1, \ldots, k\}$ and $\pi(j) \in
\{i_1+1, \ldots, i_2\}$ for all $j \in \{k-i_2+1, \ldots,
k-i_1\}$. All other constraints are satisfied with strict
inequality.
We define vector $(e_1,\ldots, e_k)$ as
\begin{align}
\left\{
\begin{array}{ll}
e_i =\frac{\epsilon}{\sum_{i=
k-i_1+1}^{k} \frac{1}{i}} & \ \textrm{for} \ i=1,\ldots, i_1 \\
e_i  =\frac{-\epsilon}{\sum_{i=
k-i_2+1}^{k-i_1} \frac{1}{i}} &  \ \textrm{for} \   i=i_1+1,\ldots, i_2\\
e_i =0 & \ \textrm{otherwise}
\end{array}
\right.
\end{align}

An explicit calculation shows that for any $\epsilon>0$,
both the point
$$ (\td_1, \ldots, \td_k) + (e_1,\ldots, e_k)$$ and the point
$$ (\td_1, \ldots, \td_k) - (e_1,\ldots, e_k)$$
 continue to satisfy the tight inequalities with equality. Moreover,
 for $\epsilon$ sufficiently small, the constraints that are not
 tight on $(\td_1, \ldots, \td_k)$ remain not tight on these $2$
 points. Hence, both these points lie in the polyhedron, and hence
 $(\td_1, \ldots, \td_k)$, which is the average between these points, cannot
be a corner point.

Thus, the only point in the strict positive quadrant that can be a
corner point of the polyhedron is the point:
$$ \frac{1}{\sum_{i=1}^k \frac{1}{i}} (1,1, \ldots, 1).$$
This point is achievable by Theorem \ref{thm:squre}. Any other point
in the polyhedron is a convex combination of this point and points
for which some of the coordinates are zero. Each one of these latter
points is in fact in the polyhedron for some smaller value of $K= k'
< k$. By the induction hypothesis, each of these points is
achievable. Hence, by time-sharing, any point in the polyhedron for
$K=k$ is achievable.

\section{Achievable Scheme for Theorem~\ref{thm:nonsqure}}~\label{Sec:non-square:original}
In Section~\ref{sec:achievable-square}, we explained  an algorithm
to achieve ${\DoF}^*_1(M, K)$, when $M \geq K$. More generally, we
characterized ${\DoF}^*_j(M, K)$, when $M \geq K-j+1$. In this
section, we extend the optimal achievable scheme of
Section~\ref{sec:achievable-square} and develop a sub-optimal
algorithm for the case that $M < K-j+1$ for order--$j$ messages. We
first focus on the case $M=2$ and $K=3$.

\subsection{Achievable Scheme for $M=2$,  $K=3$}
From Theorems~\ref{thm:squre} and \ref{thm:outer--bound--Gen}, we have ${\DoF}^*_2(2,3)=\frac{6}{5}$ and ${\DoF}^*_3(2,3)=1$. However, for order--one messages, we only know from  the outer--bound that ${\DoF}^*_1(2,3) \leq \frac{3}{2}$.   On the other hand, in terms of achievability, it is easy to see that  ${\DoF}^*_1(2,3) \geq {\DoF}^*_1(2,2)=\frac{4}{3}$ which can be achieved by simply ignoring one of the receivers. Now the question is whether ${\DoF}^*_1(2,3)$ is indeed the same as ${\DoF}^*_1(2,2)$ or the extra receiver can be exploited to achieve ${\DoF}$ beyond ${\DoF}^*_1(2,2)$. Here we propose an algorithm to show that ${\DoF}^*_1(2,3) > {\DoF}^*_1(2,2)$.

\begin{figure*}[tbhp]
\centering
\scalebox{0.7} 
{
\begin{pspicture}(0,-3.32)(26.08,3.36)
\psline[linewidth=0.04cm,arrowsize=0.05291667cm 2.0,arrowlength=1.4,arrowinset=0.4]{->}(6.62,-0.06)(7.44,-0.06)
\psline[linewidth=0.04cm,arrowsize=0.05291667cm 2.0,arrowlength=1.4,arrowinset=0.4]{->}(6.64,-1.42)(7.46,-1.42)
\usefont{T1}{ptm}{m}{n}
\rput(6.87,0.165){$X_1$}
\usefont{T1}{ptm}{m}{n}
\rput(6.93,-1.155){$X_2$}
\pspolygon[linewidth=0.04](7.608567,0.07773483)(7.914077,0.07773483)(7.761322,-0.13245316)
\psline[linewidth=0.04cm](7.761322,-0.11243526)(7.761322,-0.32262325)
\pspolygon[linewidth=0.04](7.608567,-1.1422652)(7.914077,-1.1422652)(7.761322,-1.3524531)
\psline[linewidth=0.04cm](7.761322,-1.3324353)(7.761322,-1.5426233)
\psframe[linewidth=0.04,dimen=outer](8.06,0.32)(7.46,-1.96)
\usefont{T1}{ptm}{m}{n}
\rput(0.53,0.085){$u_A$}
\psframe[linewidth=0.04,dimen=outer](0.92,0.48)(0.06,-0.34)
\psframe[linewidth=0.04,dimen=outer](0.92,-0.92)(0.04,-1.74)
\usefont{T1}{ptm}{m}{n}
\rput{90.0}(1.74,0.64000005){\rput(0.55,1.185){$m=1$}}
\usefont{T1}{ptm}{m}{n}
\rput{90.0}(2.7800002,-0.39999998){\rput(1.59,1.185){$m=2$}}
\usefont{T1}{ptm}{m}{n}
\rput{90.0}(3.94,-1.56){\rput(2.75,1.185){$m=3$}}
\usefont{T1}{ptm}{m}{n}
\rput{90.0}(4.98,-2.6){\rput(3.79,1.185){$m=4$}}
\usefont{T1}{ptm}{m}{n}
\rput{90.0}(6.06,-3.6799998){\rput(4.87,1.185){$m=5$}}
\usefont{T1}{ptm}{m}{n}
\rput{90.0}(7.1,-4.68){\rput(5.89,1.205){$m=6$}}
\usefont{T1}{ptm}{m}{n}
\rput(0.49,-1.315){$v_A$}
\usefont{T1}{ptm}{m}{n}
\rput(1.55,0.085){$w_A$}
\psframe[linewidth=0.04,dimen=outer](1.98,0.48)(1.12,-0.34)
\psframe[linewidth=0.04,dimen=outer](1.98,-0.92)(1.1,-1.74)
\usefont{T1}{ptm}{m}{n}
\rput(1.54,-1.335){$\psi_A$}
\psline[linewidth=0.04cm,arrowsize=0.05291667cm 2.0,arrowlength=1.4,arrowinset=0.4]{->}(11.74,1.22)(12.56,1.22)
\psline[linewidth=0.04cm,arrowsize=0.05291667cm 2.0,arrowlength=1.4,arrowinset=0.4]{->}(11.8,-0.78)(12.62,-0.78)
\usefont{T1}{ptm}{m}{n}
\rput(12.22,-0.475){$Y_B$}
\pspolygon[linewidth=0.04](11.308567,-0.56226516)(11.614077,-0.56226516)(11.461322,-0.7724532)
\psline[linewidth=0.04cm](11.461322,-0.75243527)(11.461322,-0.96262324)
\usefont{T1}{ptm}{m}{n}
\rput(12.17,1.525){$Y_A$}
\pspolygon[linewidth=0.04](11.268567,1.4377348)(11.574077,1.4377348)(11.421322,1.2275468)
\psline[linewidth=0.04cm](11.421322,1.2475648)(11.421322,1.0373768)
\psline[linewidth=0.03cm,arrowsize=0.05291667cm 2.0,arrowlength=1.4,arrowinset=0.4]{->}(8.16,-0.14)(11.22,1.18)
\psline[linewidth=0.03cm,arrowsize=0.05291667cm 2.0,arrowlength=1.4,arrowinset=0.4]{->}(8.12,-1.28)(11.26,1.14)
\psline[linewidth=0.03cm,arrowsize=0.05291667cm 2.0,arrowlength=1.4,arrowinset=0.4]{->}(8.16,-0.14)(11.14,-0.76)
\psline[linewidth=0.03cm,arrowsize=0.05291667cm 2.0,arrowlength=1.4,arrowinset=0.4]{->}(8.12,-1.3)(11.12,-0.8)
\psframe[linewidth=0.03,dimen=outer](8.68,2.6)(7.58,1.96)
\usefont{T1}{ptm}{m}{n}
\rput(8.13,2.285){Delay}
\psline[linewidth=0.02,arrowsize=0.05291667cm 2.0,arrowlength=1.4,arrowinset=0.4]{->}(9.86,1.12)(9.86,2.28)(8.66,2.28)
\psline[linewidth=0.04cm,arrowsize=0.05291667cm 2.0,arrowlength=1.4,arrowinset=0.4]{->}(11.82,-2.92)(12.64,-2.92)
\usefont{T1}{ptm}{m}{n}
\rput(12.18,-2.635){$Y_C$}
\pspolygon[linewidth=0.04](11.3285675,-2.7022653)(11.634077,-2.7022653)(11.481322,-2.9124532)
\psline[linewidth=0.04cm](11.481322,-2.8924353)(11.481322,-3.1026232)
\psline[linewidth=0.03cm,arrowsize=0.05291667cm 2.0,arrowlength=1.4,arrowinset=0.4]{->}(8.14,-0.14)(11.08,-2.92)
\psline[linewidth=0.03cm,arrowsize=0.05291667cm 2.0,arrowlength=1.4,arrowinset=0.4]{->}(8.12,-1.32)(11.04,-2.94)
\psellipse[linewidth=0.03,linestyle=dashed,dash=0.16cm 0.16cm,dimen=outer](9.85,-0.7)(0.15,1.84)
\psline[linewidth=0.03,arrowsize=0.05291667cm 2.0,arrowlength=1.4,arrowinset=0.4]{->}(7.58,2.3)(6.82,2.3)(6.84,0.54)
\usefont{T1}{ptm}{m}{n}
\rput(13.81,1.285){\small $L_1(u_A, v_A)$}
\psframe[linewidth=0.04,dimen=outer](14.78,1.68)(12.88,0.86)
\usefont{T1}{ptm}{m}{n}
\rput(13.89,3.125){$m=1$}
\usefont{T1}{ptm}{m}{n}
\rput(13.81,-0.895){\small $L_2(u_A, v_A)$}
\psframe[linewidth=0.04,dimen=outer](14.78,-0.5)(12.88,-1.32)
\usefont{T1}{ptm}{m}{n}
\rput(13.79,-2.895){\small $L_3(u_A, v_A)$}
\psframe[linewidth=0.04,dimen=outer](14.76,-2.5)(12.86,-3.32)
\usefont{T1}{ptm}{m}{n}
\rput(15.95,1.305){\small $L_4(w_A, \psi_A)$}
\psframe[linewidth=0.04,dimen=outer](16.92,1.7)(15.02,0.88)
\usefont{T1}{ptm}{m}{n}
\rput(16.05,3.145){$m=2$}
\psframe[linewidth=0.04,dimen=outer](16.92,-0.48)(15.02,-1.3)
\psframe[linewidth=0.04,dimen=outer](16.9,-2.48)(15.0,-3.3)
\usefont{T1}{ptm}{m}{n}
\rput(15.93,-0.895){\small $L_5(w_A, \psi_A)$}
\usefont{T1}{ptm}{m}{n}
\rput(15.95,-2.875){\small $L_6(w_A, \psi_A)$}
\usefont{T1}{ptm}{m}{n}
\rput(14.86,0.145){\small $\hat{L}_2(u_A,v_A, w_A, \psi_A)$}
\usefont{T1}{ptm}{m}{n}
\rput(14.92,-1.875){\small $\hat{L}_3(u_A,v_A, w_A, \psi_A)$}
\psline[linewidth=0.04cm,linestyle=dashed,dash=0.16cm 0.16cm,arrowsize=0.05291667cm 2.0,arrowlength=1.4,arrowinset=0.4]{->}(13.8,-0.48)(14.9,-0.12)
\psline[linewidth=0.04cm,linestyle=dashed,dash=0.16cm 0.16cm,arrowsize=0.05291667cm 2.0,arrowlength=1.4,arrowinset=0.4]{->}(15.94,-0.5)(15.0,-0.12)
\psline[linewidth=0.04cm,linestyle=dashed,dash=0.16cm 0.16cm,arrowsize=0.05291667cm 2.0,arrowlength=1.4,arrowinset=0.4]{->}(13.78,-2.5)(14.88,-2.14)
\psline[linewidth=0.04cm,linestyle=dashed,dash=0.16cm 0.16cm,arrowsize=0.05291667cm 2.0,arrowlength=1.4,arrowinset=0.4]{->}(15.96,-2.5)(15.02,-2.12)
\usefont{T1}{ptm}{m}{n}
\rput(2.72,0.085){$u_B$}
\psframe[linewidth=0.04,dimen=outer](3.12,0.48)(2.26,-0.34)
\psframe[linewidth=0.04,dimen=outer](3.12,-0.92)(2.24,-1.74)
\usefont{T1}{ptm}{m}{n}
\rput(2.68,-1.315){$v_B$}
\usefont{T1}{ptm}{m}{n}
\rput(3.74,0.085){$w_B$}
\psframe[linewidth=0.04,dimen=outer](4.18,0.48)(3.32,-0.34)
\psframe[linewidth=0.04,dimen=outer](4.18,-0.92)(3.3,-1.74)
\usefont{T1}{ptm}{m}{n}
\rput(3.73,-1.335){$\psi_B$}
\usefont{T1}{ptm}{m}{n}
\rput(4.84,0.085){$u_C$}
\psframe[linewidth=0.04,dimen=outer](5.24,0.48)(4.38,-0.34)
\psframe[linewidth=0.04,dimen=outer](5.24,-0.92)(4.36,-1.74)
\usefont{T1}{ptm}{m}{n}
\rput(4.8,-1.315){$v_C$}
\usefont{T1}{ptm}{m}{n}
\rput(5.86,0.085){$w_C$}
\psframe[linewidth=0.04,dimen=outer](6.3,0.48)(5.44,-0.34)
\psframe[linewidth=0.04,dimen=outer](6.3,-0.92)(5.42,-1.74)
\usefont{T1}{ptm}{m}{n}
\rput(5.85,-1.315){$\psi_C$}
\usefont{T1}{ptm}{m}{n}
\rput(18.03,1.305){\small $L_7(u_B, v_B)$}
\psframe[linewidth=0.04,dimen=outer](19.02,1.7)(17.12,0.88)
\usefont{T1}{ptm}{m}{n}
\rput(18.17,3.145){$m=3$}
\usefont{T1}{ptm}{m}{n}
\rput(18.03,-0.875){\small $L_8(u_B, v_B)$}
\psframe[linewidth=0.04,dimen=outer](19.02,-0.48)(17.12,-1.3)
\usefont{T1}{ptm}{m}{n}
\rput(18.01,-2.855){\small $L_9(u_B, v_B)$}
\psframe[linewidth=0.04,dimen=outer](19.0,-2.48)(17.1,-3.3)
\usefont{T1}{ptm}{m}{n}
\rput(20.16,1.325){\small $L_{10}(w_B, \psi_B)$}
\psframe[linewidth=0.04,dimen=outer](21.16,1.72)(19.26,0.9)
\usefont{T1}{ptm}{m}{n}
\rput(20.29,3.145){$m=4$}
\psframe[linewidth=0.04,dimen=outer](21.16,-0.46)(19.26,-1.28)
\psframe[linewidth=0.04,dimen=outer](21.14,-2.46)(19.24,-3.28)
\usefont{T1}{ptm}{m}{n}
\rput(20.14,-0.875){\small $L_{11}(w_B, \psi_B)$}
\usefont{T1}{ptm}{m}{n}
\rput(20.12,-2.875){\small $L_{12}(w_B, \psi_B)$}
\usefont{T1}{ptm}{m}{n}
\rput(19.18,2.385){\small $\hat{L}_4(u_B,v_B, w_B, \psi_B)$}
\usefont{T1}{ptm}{m}{n}
\rput(19.12,-1.855){\small $\hat{L}_6(u_B,v_B, w_B, \psi_B)$}
\psline[linewidth=0.04cm,linestyle=dashed,dash=0.16cm 0.16cm,arrowsize=0.05291667cm 2.0,arrowlength=1.4,arrowinset=0.4]{->}(17.98,1.72)(19.08,2.08)
\psline[linewidth=0.04cm,linestyle=dashed,dash=0.16cm 0.16cm,arrowsize=0.05291667cm 2.0,arrowlength=1.4,arrowinset=0.4]{->}(20.3,1.72)(19.36,2.1)
\psline[linewidth=0.04cm,linestyle=dashed,dash=0.16cm 0.16cm,arrowsize=0.05291667cm 2.0,arrowlength=1.4,arrowinset=0.4]{->}(18.02,-2.48)(19.12,-2.12)
\psline[linewidth=0.04cm,linestyle=dashed,dash=0.16cm 0.16cm,arrowsize=0.05291667cm 2.0,arrowlength=1.4,arrowinset=0.4]{->}(20.2,-2.48)(19.26,-2.1)
\usefont{T1}{ptm}{m}{n}
\rput(22.24,1.305){\small $L_{13}(u_C, v_C)$}
\psframe[linewidth=0.04,dimen=outer](23.24,1.72)(21.34,0.9)
\usefont{T1}{ptm}{m}{n}
\rput(22.33,3.165){$m=5$}
\usefont{T1}{ptm}{m}{n}
\rput(22.24,-0.855){\small $L_{14}(u_C, v_C)$}
\psframe[linewidth=0.04,dimen=outer](23.24,-0.46)(21.34,-1.28)
\usefont{T1}{ptm}{m}{n}
\rput(22.24,-2.855){\small $L_{15}(u_C, v_C)$}
\psframe[linewidth=0.04,dimen=outer](23.22,-2.46)(21.32,-3.28)
\usefont{T1}{ptm}{m}{n}
\rput(24.34,1.345){\small $L_{16}(w_C, \psi_C)$}
\psframe[linewidth=0.04,dimen=outer](25.38,1.74)(23.48,0.92)
\usefont{T1}{ptm}{m}{n}
\rput(24.43,3.165){$m=6$}
\psframe[linewidth=0.04,dimen=outer](25.38,-0.44)(23.48,-1.26)
\psframe[linewidth=0.04,dimen=outer](25.36,-2.44)(23.46,-3.26)
\usefont{T1}{ptm}{m}{n}
\rput(24.36,-0.835){\small $L_{17}(w_C, \psi_C)$}
\usefont{T1}{ptm}{m}{n}
\rput(24.34,-2.855){\small $L_{18}(w_C, \psi_C)$}
\usefont{T1}{ptm}{m}{n}
\rput(23.4,2.405){\small $\hat{L}_7(u_C,v_C, w_C, \psi_C)$}
\usefont{T1}{ptm}{m}{n}
\rput(23.46,0.245){\small $\hat{L}_8(u_C,v_C, w_C, \psi_C)$}
\psline[linewidth=0.04cm,linestyle=dashed,dash=0.16cm 0.16cm,arrowsize=0.05291667cm 2.0,arrowlength=1.4,arrowinset=0.4]{->}(22.2,1.74)(23.3,2.1)
\psline[linewidth=0.04cm,linestyle=dashed,dash=0.16cm 0.16cm,arrowsize=0.05291667cm 2.0,arrowlength=1.4,arrowinset=0.4]{->}(24.52,1.74)(23.58,2.12)
\psline[linewidth=0.04cm,linestyle=dashed,dash=0.16cm 0.16cm,arrowsize=0.05291667cm 2.0,arrowlength=1.4,arrowinset=0.4]{->}(22.24,-0.46)(23.34,-0.1)
\psline[linewidth=0.04cm,linestyle=dashed,dash=0.16cm 0.16cm,arrowsize=0.05291667cm 2.0,arrowlength=1.4,arrowinset=0.4]{->}(24.54,-0.44)(23.6,-0.06)
\end{pspicture} 
}

\centering
\caption{A Sub-Optimal Scheme for $M=2$ and $K=3$, The First Phase}
\label{fig:feedback_23_Sub_Opt}
\end{figure*}
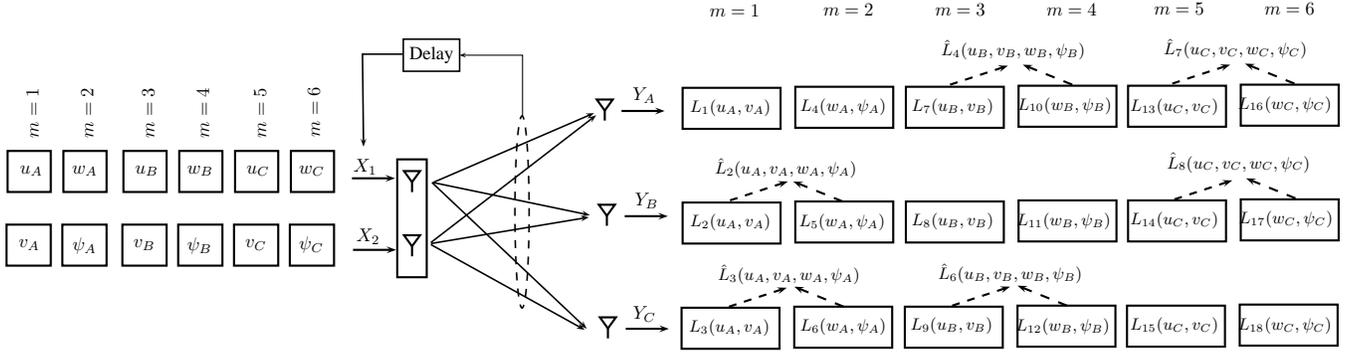

The achievable scheme is as follows. Let $u_r$,  $v_r$, $w_r$, and $\psi_r$ be four symbols for receiver $r$, $r=A, B, C$.  The first phase of the scheme has 6 time--slots. The first two time--slots are dedicated to receiver $A$. In these two time--slots, the transmitter sends four random linear combinations of $u_A$,  $v_A$, $w_A$, and $\psi_A$ through the two transmit antennas. As a particular example, in the first time slot, the transmitter sends $u_A$ and $v_A$, and in the second time slot, it sends $w_A$ and $\psi_A$. Refer to Fig.~\ref{fig:feedback_23_Sub_Opt} for details. Similarly,  in time--slots 3 and 4, the transmitter sends four random linear combinations of $u_B$,  $v_B$, $w_B$, and $\psi_B$.  In time--slots 5 and 6, the transmitter sends four random linear combinations of $u_C$,  $v_C$, $w_C$, and $\psi_C$.

Referring to Fig.~\ref{fig:feedback_23_Sub_Opt},   we have the following observations:

\begin{itemize}
\item Receiver $A$ already has two independent linear equations $L_1(u_A, v_A)$ and $L_4(w_A, \psi_A)$ of $u_A$,  $v_A$, $w_A$, and $\psi_A$. Therefore, it needs two more equations.
\item The four overheard equations in  $L_2(u_A, v_A)$, $L_3(u_A, v_A)$, $L_5(w_A, \psi_A)$, and $L_6(w_A, \psi_A)$  are {\emph not } linearly independent from what receiver $A$ has already received, i.e. $L_1(u_A, v_A)$ and $L_4(w_A, \psi_A)$.
\item We can purify the four overheard equations and form two equations that are linearly independent with $L_1(u_A, v_A)$ and $L_4(w_A, \psi_A)$. For example, receiver $B$ can form $\hat{L}_2(u_A, v_A, w_A, \psi_A)$ as a random linear combination of  $L_2(u_A, v_A)$ and $L_5(w_A, \psi_A)$. Similarly, receiver $C$ can form $\hat{L}_3(u_A, v_A, w_A, \psi_A)$ as a random linear combination of  $L_3(u_A, v_A)$ and $L_6(w_A, \psi_A)$. The coefficients of these linear combinations have been preselected and shared among all nodes.

\item It is easy to see that almost surely,  $\hat{L}_2(u_A, v_A, w_A, \psi_A)$ and $\hat{L}_3(u_A, v_A, w_A, \psi_A)$ are linearly independent of $L_1(u_A, v_A)$ and $L_4(w_A, \psi_A)$.

\item If somehow  we deliver $\hat{L}_2(u_A, v_A, w_A, \psi_A)$   and $\hat{L}_3(u_A, v_A, w_A, \psi_A)$  to receiver $A$, then it has enough equations to solve for $u_A$,  $v_A$, $w_A$, and $\psi_A$.

\end{itemize}

Similarly, as shown in Fig.~\ref{fig:feedback_23_Sub_Opt}, we can purify the overheard equations in time--slots dedicated to receivers $B$ and $C$.  Now, the available side information and the requirements are the same as those we had after  phase one for the case of $M=K=3$ (see Subsection~\ref{subsec:M=K=3}). Equations  $\hat{L}_2(u_A, v_A, w_A, \psi_A)$   and $\hat{L}_3(u_A, v_A, w_A, \psi_A)$ are available at receivers $B$ and $C$, respectively,  and are needed by receiver $A$,   equations  $\hat{L}_4(u_B, v_B, w_B, \psi_B)$   and $\hat{L}_6(u_B, v_B, w_B, \psi_B)$ are available at receivers $A$ and $C$, respectively,  and are needed by receiver $B$,  and equations $\hat{L}_7(u_C, v_C, w_C, \psi_C)$   and $\hat{L}_8(u_C, v_C, w_C, \psi_C)$ are available at receivers $A$ and $B$, respectively,  and are needed by receiver $C$. We define
\begin{align}
u_{AB}= \hat{L}_2(u_A, v_A, w_A, \psi_A)+ \hat{L}_4(u_B, v_B, w_B, \psi_B),\\
u_{AC}=\hat{L}_3(u_A, v_A, w_A, \psi_A)+ \hat{L}_7(u_C, v_C, w_C, \psi_C),\\
u_{BC}= \hat{L}_6(u_B, v_B, w_B, \psi_B)+\hat{L}_8(u_C, v_C, w_C, \psi_C).
\end{align}

Considering the available overheard equations at each receiver,  one can easily conclude that $u_{AB}$ is needed by both receivers $A$ and $B$,  $u_{AC}$ is needed by both receivers $A$ and $C$, and $u_{BC}$ is needed by both receivers $B$ and $C$.  The transmitter needs $\frac{3}{{\DoF}_2^*(2,3)}$ time--slots to deliver these three order--two symbols, where according to Theorem~\ref{thm:squre}, ${\DoF}_2^*(2,3)=\frac{6}{5}$.  In summary, phase one starts with 12 order--one messages, takes 6 time--slots, and generates 3 order--two symbols. Therefore, we achieve
\begin{align}\label{DoF_1(2,3)}
{\DoF}_1(2,3)=\frac{12}{6+ \frac{3}{{\DoF}^*_2(2,3)}}=\frac{24}{17},
\end{align}
which is strictly greater than ${\DoF}^*_1(2,2)=\frac{4}{3}$. Therefore, the proposed achievable scheme exploits  the extra receiver to improve ${\DoF}$.  However, we notice that the achieved ${\DoF}_1(2,3)$ of $\frac{24}{17}$ is still less than $\frac{3}{2}=\frac{24}{16}$ which is suggested by the outer--bound.

%
%

\subsection{General Proof for Theorem~\ref{thm:nonsqure}}

Here, we explain a general version of the proposed algorithm.  Again the algorithm includes $K-j+1$ phases.  Phase $j$ takes symbols of order $j$ (meaning that it is needed by $j$ receivers simultaneously),  and generates symbols of  order $j+1$. For $j=K$, the phase is simple and generates no more symbols.

Let us define $q_j$ as
\begin{align}
q_j= \min \{M-1, K-j\}.
\end{align}
In addition, we define $\eta_j$ as the greatest common factor of $q_j$ and $K-j$, i.e.
\begin{align}
\eta_j= \textrm{gcf} \{q_j, K-j\}.
\end{align}

Phase $j$ takes $(K-j) \frac{q_j+1}{\eta_j} {K \choose j}$ symbols of order $j$ and yields $j \frac{q_j}{\eta_j}{K \choose {j+1}}$ symbols with order $j+1$. This phase has ${K \choose j}$ sub-phases, where each sub-phase is dedicated to a subset $\mathcal{S}$ of the receivers, $|\mathcal{S}|=j$.  The sub-phase dedicated to subset $\mathcal{S}$ is   denoted by $\textrm{S-Ph}(\mathcal{S})$. Each sub-phase takes $\frac{K-j}{\eta_j}$ time-slots. In $\textrm{S-Ph}(\mathcal{S})$, the transmitter sends random linear combinations of $\beta_j= \frac{(q_j +1)(K-j)}{\eta_j}$ symbols $u_{\mathcal{S},1}, u_{\mathcal{S},2}, \ldots, u_{\mathcal{S}, \beta_j}$, desired by all receivers in $\mathcal{S}$. The transmitter uses at least $q_j+1$ of the transmit antennas. The linear equation of the transmitted symbols received by receiver $r$, in the $t$-th time slot of $\textrm{S-Ph}(\mathcal{S})$,  is denoted by $L_{\mathcal{\mathcal{S}},r}(t)$. Let us focus on the equations of the transmitted symbols received by all receivers in $\textrm{S-Ph}(\mathcal{S})$.  We have the following observations:
\begin{itemize}
\item For every $r$,  $r \in \mathcal{S}$, and $t$,  $t \in \{1,2, \ldots, \frac{K-j}{\eta_j} \}$, the $K-j+1$ equations  $\{L_{\mathcal{\mathcal{S}}, r'}(t)$, $r' \in \{r\} \cup \mathcal{E} \backslash \mathcal{S}  \}$ are {\emph not} necessarily linearly independent. The reason is that $|\{r\} \cup \mathcal{E} \backslash \mathcal{S}| =K-j+1$, while the number of transmit antennas is $M$ which can be less than $K-j+1$. Indeed, among the $K-j$  overheard equations $L_{\mathcal{\mathcal{S}}, r'}(t)$, $r' \in \mathcal{E} \backslash \mathcal{S} $, we can only form $q_j$ overheard equations that are simultaneously useful to receiver $r$, for any $r$ in $\mathcal{S}$. Therefore, among $\frac{(K-j)^2}{\eta_j}$  overheard equations in $\textrm{S-Ph}(\mathcal{S})$, we can form only $ \frac{q_j(K-j)}{\eta_j}$ overheard equations that are useful for any receiver $r$,  $r \in \mathcal{S}$.

\item We purify the  overheard linear combinations. To this end, receiver $r'$, $r' \in \mathcal{E} \backslash \mathcal{S}$, forms  $\frac{q_j}{\eta_j}$ linear combinations of $L_{\mathcal{\mathcal{S}}, r'}(t)$, $t=1,\ldots, \frac{K-j}{\eta_j}$. The resultant equations are denoted by $\hat{L}_{ \mathcal{\mathcal{S}}, r'}(i), \hat{L}_{\mathcal{\mathcal{S}}, r'}(2), \ldots, \hat{L}_{\mathcal{\mathcal{S}}, r'}\left(\frac{q_j}{\eta_j}\right)$. The coefficients of the linear combinations have been preselected and shared among all nodes.
It is easy to see that for every $r$, the following $\frac{(q_j+1)(K-j)}{\eta_j}$ equations are linearly independent:  $L_{ \mathcal{\mathcal{S}},r}(t)$, $t=1, \ldots, \frac{K-j}{\eta_j}$,  and  $\hat{L}_{ \mathcal{\mathcal{S}}, r'}(\hat{t})$, $r' \in \mathcal{E} \backslash \mathcal{S} $ and $\hat{t} \in 1,\ldots, \frac{q_j}{\eta_j}$. Therefore, if we somehow deliver $\hat{L}_{\mathcal{\mathcal{S}},r'}(\hat{t})$, $r' \in \mathcal{E} \backslash \mathcal{S} $ and $\hat{t} \in 1,\ldots, \frac{q_j}{\eta_j}$ to receiver $r$, $r \in \mathcal{S}$, then it will have $\beta_j=\frac{(q_j+1)(K-j)}{ \eta_j}$ linearly independent equations to solve for all desired symbols $u_{\mathcal{S},1}, u_{\mathcal{S},2}, \ldots, u_{\mathcal{S},\beta_j}$.

\item Having the above two observations, we note that the purified  linear combinations by receiver $r'$,  $r' \in \mathcal{E} \backslash \mathcal{S}$, are simultaneously useful for all receivers in $\mathcal{S}$.
\end{itemize}

After repeating the above transmission for all $\mathcal{S}$, where $\mathcal{S}\subset \mathcal{E}$ and $|\mathcal{S}|=j$, then we have another important property.  Consider a subset $\T$ of the receivers, where $|\T|=j+1$. Then each receiver $r$, $r \in \T$, has $\frac{q_j}{\eta_j}$ purified  linear combination $\hat{L}_{\T \backslash \{ r \}, r} (t)$, $t=1, \ldots, \frac{q_j}{\eta_j}$, which are simultaneously useful for all receivers in $\T \backslash \{ r \}$. We note that the transmitter is aware of these purified  equations through delayed CSIT. For every $\T \subset \mathcal{E}$, $|\T|=j+1$, the transmitter forms $j \frac{q_j}{\eta_j}$ random linear combinations of $\hat{L}_{\T \backslash \{ r \}, r }(t)$, $r \in  \T$, $t=1, \ldots, \frac{q_j}{\eta_j}$,  denoted by $u_{\T, 1},  u_{\T, 2}, \ldots, u_{\T, j\frac{q_j}{\eta_j}}$. We note that $u_{\T, \xi}$, $1 \leq  \xi \leq j\frac{q_j}{\eta_j}$, is simultaneously useful for all receivers in $\T$. The reason is that each receiver $r$,  $r  \in \T$, can subtract the contributions of $\hat{L}_{\T \backslash \{ r \}, r}(t)$, $t=1, \ldots, \frac{q_j}{\eta_j}$, from $u_{\T, \xi}$, $\xi = 1, \ldots, j\frac{q_j}{\eta_j}$, and form $j\frac{q_j}{\eta_j}$ linearly independent combinations of $\hat{L}_{\T \backslash \{ r' \},r' }(t)$, $r' \in  \T \backslash \{ r \}$, $t=1,\ldots, \frac{q_j}{\eta_j}$. Therefore, using the above procedure, the transmitter forms $j \frac{q_j}{\eta_j} {K \choose {j+1}}$ symbols with order $j+1$. The important observation is that if these $j  \frac{q_j}{\eta_j} {K \choose {j+1}}$ symbols are delivered to the designated receivers, then each receiver will have enough equations to solve for all designated messages with order $j$.

In summary, this phase takes  $(K-j) \frac{q_j+1}{\eta_j} {K \choose j}$ symbols of order $j$, takes  $ \frac{K-j}{\eta_j} {K \choose j}$   time--slots, and yields $j \frac{q_j}{\eta_j}{K \choose {j+1}}$ symbols of order $j+1$.
 If we have a scheme which achieves ${\DoF}_{j+1}(M,K)$ for order--$(j+1)$ symbols, then we achieve ${\DoF}_j(M,K)$,
\begin{align}
{\DoF}_j(M,K)=\frac{   (K-j) \frac{q_j+1}{\eta_j} {K \choose j} }{ \frac{K-j}{\eta_j} {K \choose j} +  \frac{j \frac{q_j}{\eta_j}{K \choose {j+1}}}{{\DoF}_{j+1}(M,K)} },
\end{align}
or
\begin{align}
\frac{ q_j+1 }{j} \frac{1}{{\DoF}_j(M,K)}=\frac{1}{j} +  \frac{q_j}{j+1 }\frac{1}{{\DoF}_{j+1}(M,K)}.
\end{align}

\section{Improved Scheme for $M=2$}\label{Sec:Imoroved_2Tr}
Recall that the scheme of Section~\ref{Sec:non-square:original} achieves ${\DoF}_1(2,3)$ of $\frac{24}{17}$. The achieved ${\DoF}$ is greater that ${\DoF}^*_1(2,2)=\frac{4}{3}$, which shows that we could exploit the extra receiver with respect to the number of transmit antennas. However, it is still smaller than $\frac{3}{2}$ which is suggested by the outer--bound. Now the question is whether the achievable scheme or the outer--bound is loose.

In what follows, we show that for $M=2$ and $K=3$, the outer--bound is tight and the achievable scheme of Section~\ref{Sec:non-square:original} is loose. Before that, we explain an alternative solution for a system with $M=K=2$. The idea of the alternative solution is the key to achieve the optimal ${\DoF}$ for the systems with $M=2$ and $K=3$.

\subsection{Alternative Scheme for $M=K=2$}\label{Sub-Sec:NewM=K=2}
\begin{figure*}[tbhp]
\centering
\scalebox{0.7} 
{
\begin{pspicture}(0,-2.61)(24.28,2.61)
\definecolor{color8455}{rgb}{0.2,0.2,0.2}
\psline[linewidth=0.04cm,arrowsize=0.05291667cm 2.0,arrowlength=1.4,arrowinset=0.4]{->}(7.04,0.51)(7.86,0.51)
\psline[linewidth=0.04cm,arrowsize=0.05291667cm 2.0,arrowlength=1.4,arrowinset=0.4]{->}(7.04,-0.71)(7.86,-0.71)
\usefont{T1}{ptm}{m}{n}
\rput(7.29,0.735){$X_1$}
\usefont{T1}{ptm}{m}{n}
\rput(7.33,-0.445){$X_2$}
\pspolygon[linewidth=0.04](8.028567,0.6477348)(8.334077,0.6477348)(8.181322,0.43754685)
\psline[linewidth=0.04cm](8.181322,0.45756474)(8.181322,0.24737674)
\pspolygon[linewidth=0.04](8.048567,-0.37226516)(8.354076,-0.37226516)(8.201322,-0.5824532)
\psline[linewidth=0.04cm](8.201322,-0.56243527)(8.201322,-0.77262324)
\psframe[linewidth=0.04,dimen=outer](8.48,0.89)(7.88,-0.97)
\psline[linewidth=0.04cm,arrowsize=0.05291667cm 2.0,arrowlength=1.4,arrowinset=0.4]{->}(12.12,1.23)(12.94,1.23)
\psline[linewidth=0.04cm,arrowsize=0.05291667cm 2.0,arrowlength=1.4,arrowinset=0.4]{->}(12.2,-1.11)(13.02,-1.11)
\usefont{T1}{ptm}{m}{n}
\rput(12.55,1.535){$Y_A$}
\usefont{T1}{ptm}{m}{n}
\rput(12.62,-0.805){$Y_B$}
\pspolygon[linewidth=0.04](11.648567,1.4477348)(11.954077,1.4477348)(11.801322,1.2375468)
\psline[linewidth=0.04cm](11.801322,1.2575648)(11.801322,1.0473768)
\pspolygon[linewidth=0.04](11.708567,-0.89226514)(12.014077,-0.89226514)(11.861322,-1.1024531)
\psline[linewidth=0.04cm](11.861322,-1.0824353)(11.861322,-1.2926233)
\psline[linewidth=0.04cm,arrowsize=0.05291667cm 2.0,arrowlength=1.4,arrowinset=0.4]{->}(8.54,0.47)(11.6,1.19)
\psline[linewidth=0.04cm,arrowsize=0.05291667cm 2.0,arrowlength=1.4,arrowinset=0.4]{->}(8.46,-0.51)(11.64,1.15)
\psline[linewidth=0.04cm,arrowsize=0.05291667cm 2.0,arrowlength=1.4,arrowinset=0.4]{->}(8.52,0.45)(11.56,-1.01)
\psline[linewidth=0.04cm,arrowsize=0.05291667cm 2.0,arrowlength=1.4,arrowinset=0.4]{->}(8.48,-0.53)(11.6,-1.11)
\psbezier[linewidth=0.04,linecolor=color8455,linestyle=dashed,dash=0.16cm 0.16cm,arrowsize=0.05291667cm 2.0,arrowlength=1.4,arrowinset=0.4]{->}(9.96,0.23)(10.7,0.19)(10.32,1.07)(9.92,1.03)
\psbezier[linewidth=0.04,linestyle=dashed,dash=0.16cm 0.16cm,arrowsize=0.05291667cm 2.0,arrowlength=1.4,arrowinset=0.4]{<-}(9.96,-1.03)(10.64,-0.83)(10.26,-0.17)(9.86,-0.21)
\usefont{T1}{ptm}{m}{n}
\rput(10.03,1.375){$\mathbf{h}_A[m]$}
\usefont{T1}{ptm}{m}{n}
\rput(10.1,-1.325){$\mathbf{h}_{B}[m]$}
\psframe[linewidth=0.04,dimen=outer](9.06,2.61)(7.96,1.97)
\usefont{T1}{ptm}{m}{n}
\rput(8.51,2.295){Delay}
\psframe[linewidth=0.04,dimen=outer](9.12,-1.97)(8.02,-2.61)
\usefont{T1}{ptm}{m}{n}
\rput(8.57,-2.305){Delay}
\psline[linewidth=0.04,arrowsize=0.05291667cm 2.0,arrowlength=1.4,arrowinset=0.4]{->}(8.04,-2.29)(7.06,-2.29)(7.06,-1.03)
\psline[linewidth=0.04,arrowsize=0.05291667cm 2.0,arrowlength=1.4,arrowinset=0.4]{->}(10.04,1.65)(10.04,2.29)(9.04,2.29)
\psline[linewidth=0.04,arrowsize=0.05291667cm 2.0,arrowlength=1.4,arrowinset=0.4]{->}(7.96,2.29)(7.1,2.29)(7.1150875,1.19)
\psline[linewidth=0.04,arrowsize=0.05291667cm 2.0,arrowlength=1.4,arrowinset=0.4]{->}(10.08,-1.71)(10.08,-2.31)(9.08,-2.31)
\usefont{T1}{ptm}{m}{n}
\rput(1.07,0.675){$u_A+u_B$}
\usefont{T1}{ptm}{m}{n}
\rput(1.25,1.595){\small $m=1$}
\psframe[linewidth=0.02,dimen=outer](2.1,1.11)(0.02,0.21)
\usefont{T1}{ptm}{m}{n}
\rput(1.03,-0.685){$v_A+v_B$}
\psframe[linewidth=0.02,dimen=outer](2.12,-0.25)(0.0,-1.15)
\usefont{T1}{ptm}{m}{n}
\rput(5.63,1.555){\small $m=3$}
\usefont{T1}{ptm}{m}{n}
\rput(3.46,0.695){$L_2(u_A,v_A)$}
\psframe[linewidth=0.02,dimen=outer](4.34,1.11)(2.36,0.21)
\usefont{T1}{ptm}{m}{n}
\rput(3.43,-0.705){$0$}
\psframe[linewidth=0.02,dimen=outer](4.34,-0.25)(2.36,-1.15)
\usefont{T1}{ptm}{m}{n}
\rput(3.61,1.595){\small $m=2$}
\usefont{T1}{ptm}{m}{n}
\rput(5.58,0.675){$L_3(u_B,v_B)$}
\psframe[linewidth=0.02,dimen=outer](6.54,1.09)(4.54,0.19)
\usefont{T1}{ptm}{m}{n}
\rput(5.59,-0.725){$0$}
\psframe[linewidth=0.02,dimen=outer](6.54,-0.27)(4.52,-1.17)
\usefont{T1}{ptm}{m}{n}
\rput(15.45,2.135){\small $m=1$}
\usefont{T1}{ptm}{m}{n}
\rput(15.36,1.295){$L_1(u_A, v_A)+L_3(u_B, v_B)$}
\psframe[linewidth=0.02,dimen=outer](17.36,1.69)(13.32,0.77)
\usefont{T1}{ptm}{m}{n}
\rput(15.34,-1.085){$L_2(u_A, v_A)+L_4(u_B, v_B)$}
\psframe[linewidth=0.02,dimen=outer](17.34,-0.67)(13.3,-1.59)
\usefont{T1}{ptm}{m}{n}
\rput(19.11,1.255){$h_{A1}[2]L_2(u_A,v_A)$}
\psframe[linewidth=0.02,dimen=outer](20.62,1.69)(17.72,0.77)
\usefont{T1}{ptm}{m}{n}
\rput(19.21,2.055){\small $m=2$}
\usefont{T1}{ptm}{m}{n}
\rput(19.12,-1.125){$h_{B1}[2]L_2(u_A,v_A)$}
\psframe[linewidth=0.02,dimen=outer](20.62,-0.69)(17.74,-1.59)
\usefont{T1}{ptm}{m}{n}
\rput(22.17,1.275){$h_{A1}[3]L_3(u_B,v_B)$}
\psframe[linewidth=0.02,dimen=outer](23.64,1.69)(20.82,0.83)
\usefont{T1}{ptm}{m}{n}
\rput(22.31,2.095){\small $m=3$}
\usefont{T1}{ptm}{m}{n}
\rput(22.22,-1.105){$h_{B1}[3]L_3(u_B,v_B)$}
\psframe[linewidth=0.02,dimen=outer](23.64,-0.69)(20.84,-1.61)
\end{pspicture} 
}
\centering
\caption{Alternative Achievable Scheme for $M=K=2$}
\label{fig:feedback_22_Alt}
\end{figure*}
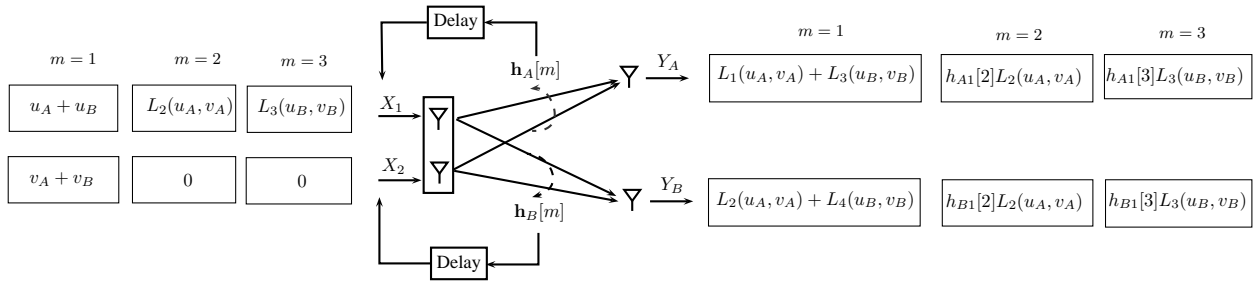

Phase one of the algorithm takes order--one messages. Let us assume that the transmitter has $u_A$ and  $v_A$ for receiver $A$ and  $u_B$ and  $v_B$ for receiver $B$. Here, phase one takes only one time--slot which is dedicated to both receivers. In this time--slot, the transmitter sends random linear combinations of all four symbols $u_A$ and  $v_A$,  $u_B$,  and $v_B$.  Refer to Fig.~\ref{fig:feedback_22_Alt} to see the details of particular examples for the linear combinations.  Receiver $A$ receives a linear combination of all four symbols.
We denote this linear combination by $L_1(u_A, v_A)+L_3(u_B,v_B)$, where $L_1(u_A, v_A)$ represents the contribution of $u_A$ and $v_A$, and $L_3(u_B, v_B)$ represents the contribution of $u_B$ and $v_B$.  Similarly, receiver $B$ receives a linear combination of all four symbols denoted by $L_2(u_A, v_A)+L_4(u_B,v_B)$.

Then, we have the following observations:
\begin{itemize}
\item If we somehow give  $L_3(u_B,v_B)$ to receiver $A$, then receiver $A$ can compute $L_1(u_A, v_A)$ by subtracting  $L_3(u_B,v_B)$ from what it already has. Then if we also give  $L_2(u_A, v_A)$ to receiver $A$, then it has two equations to solve for $u_A$ and $v_A$.

\item If we somehow give  $L_2(u_A,v_A)$ to receiver $B$, then receiver $B$ can compute $L_4(u_B, v_B)$ by subtracting  $L_2(u_A,v_A)$ from what it already has. Then if we also give  $L_3(u_B, v_B)$ to receiver $B$, then it has two equations to solve for $u_B$ and $v_B$.
\end{itemize}
In other words, both receivers $A$ and $B$ want  $L_2(u_A,v_A)$ and $L_3(u_B,v_B)$. Therefore, we can define two order--two symbols $u_{AB}$  and  $v_{AB}$ as
\begin{align}
u_{AB}= L_2(u_A,v_A),\\
v_{AB}= L_3(u_B,v_B).
\end{align}

In summary, this phase starts with 4 order--one symbols, takes one time-slot, and provides two order--two symbols. Two order--two symbols take $\frac{2}{{\DoF}_2^*(2,2)}$ time--slots to deliver.  Therefore, we achieve
\begin{align}\label{DoF_1(2,2)_alt}
{\DoF}_1(2,2)=\frac{ 4 }{  1 +  \frac{2}{{\DoF}^*_{2}(2,2)}}.
\end{align}
Since  ${\DoF}^*_{2}(2,2)=1$, this scheme achieves ${\DoF}^*_1(2,2)=\frac{4}{3}$.

\subsection{Optimal Scheme for $M=2$ and $K=3$}\label{subsec:alt-2-3}
Here, we explain an algorithm for the systems with $M=2$ and $K=3$. The first phase of this algorithm takes 12 order--one messages, takes 3 time--slots, and gives 6 order--two symbols. This sub-algorithm leads to an optimal scheme for systems with $M=2$ and $K=3$.

\begin{figure*}[tbhp]
\centering
\scalebox{0.7} 
{
\begin{pspicture}(0,-2.98)(25.4,2.98)
\psline[linewidth=0.04cm,arrowsize=0.05291667cm 2.0,arrowlength=1.4,arrowinset=0.4]{->}(5.68,0.32)(6.5,0.32)
\psline[linewidth=0.04cm,arrowsize=0.05291667cm 2.0,arrowlength=1.4,arrowinset=0.4]{->}(5.7,-1.04)(6.52,-1.04)
\usefont{T1}{ptm}{m}{n}
\rput(5.93,0.545){$X_1$}
\usefont{T1}{ptm}{m}{n}
\rput(5.99,-0.775){$X_2$}
\pspolygon[linewidth=0.04](6.668567,0.45773482)(6.9740767,0.45773482)(6.821322,0.24754684)
\psline[linewidth=0.04cm](6.821322,0.26756474)(6.821322,0.057376746)
\pspolygon[linewidth=0.04](6.668567,-0.76226515)(6.9740767,-0.76226515)(6.821322,-0.9724532)
\psline[linewidth=0.04cm](6.821322,-0.95243526)(6.821322,-1.1626233)
\psframe[linewidth=0.04,dimen=outer](7.12,0.7)(6.52,-1.58)
\usefont{T1}{ptm}{m}{n}
\rput(0.89,0.365){$u_A+u_B$}
\psframe[linewidth=0.04,dimen=outer](1.66,0.8)(0.1,-0.1)
\usefont{T1}{ptm}{m}{n}
\rput(0.91,-0.955){$v_A+v_B$}
\psframe[linewidth=0.04,dimen=outer](1.66,-0.5)(0.12,-1.4)
\usefont{T1}{ptm}{m}{n}
\rput(2.67,0.365){$w_A+u_C$}
\psframe[linewidth=0.04,dimen=outer](3.46,0.8)(1.9,-0.1)
\usefont{T1}{ptm}{m}{n}
\rput(2.7,-0.915){$\psi_A+v_C$}
\psframe[linewidth=0.04,dimen=outer](3.46,-0.5)(1.92,-1.4)
\usefont{T1}{ptm}{m}{n}
\rput(4.38,0.365){$w_B+w_C$}
\psframe[linewidth=0.04,dimen=outer](5.18,0.8)(3.62,-0.1)
\usefont{T1}{ptm}{m}{n}
\rput(4.38,-0.955){$\psi_B+\psi_C$}
\psframe[linewidth=0.04,dimen=outer](5.18,-0.5)(3.64,-1.4)
\psline[linewidth=0.04cm,arrowsize=0.05291667cm 2.0,arrowlength=1.4,arrowinset=0.4]{->}(10.8,1.6)(11.62,1.6)
\psline[linewidth=0.04cm,arrowsize=0.05291667cm 2.0,arrowlength=1.4,arrowinset=0.4]{->}(10.86,-0.4)(11.68,-0.4)
\usefont{T1}{ptm}{m}{n}
\rput(11.28,-0.095){$Y_B$}
\pspolygon[linewidth=0.04](10.368567,-0.18226518)(10.674077,-0.18226518)(10.521322,-0.39245316)
\psline[linewidth=0.04cm](10.521322,-0.37243527)(10.521322,-0.58262324)
\usefont{T1}{ptm}{m}{n}
\rput(11.23,1.905){$Y_A$}
\pspolygon[linewidth=0.04](10.3285675,1.8177348)(10.634077,1.8177348)(10.481322,1.6075468)
\psline[linewidth=0.04cm](10.481322,1.6275648)(10.481322,1.4173768)
\psline[linewidth=0.03cm,arrowsize=0.05291667cm 2.0,arrowlength=1.4,arrowinset=0.4]{->}(7.22,0.24)(10.28,1.56)
\psline[linewidth=0.03cm,arrowsize=0.05291667cm 2.0,arrowlength=1.4,arrowinset=0.4]{->}(7.18,-0.9)(10.32,1.52)
\psline[linewidth=0.03cm,arrowsize=0.05291667cm 2.0,arrowlength=1.4,arrowinset=0.4]{->}(7.22,0.24)(10.2,-0.38)
\psline[linewidth=0.03cm,arrowsize=0.05291667cm 2.0,arrowlength=1.4,arrowinset=0.4]{->}(7.18,-0.92)(10.18,-0.42)
\psframe[linewidth=0.03,dimen=outer](7.74,2.98)(6.64,2.34)
\usefont{T1}{ptm}{m}{n}
\rput(7.19,2.665){Delay}
\psline[linewidth=0.02,arrowsize=0.05291667cm 2.0,arrowlength=1.4,arrowinset=0.4]{->}(8.92,1.5)(8.92,2.66)(7.72,2.66)
\psline[linewidth=0.04cm,arrowsize=0.05291667cm 2.0,arrowlength=1.4,arrowinset=0.4]{->}(10.88,-2.54)(11.7,-2.54)
\usefont{T1}{ptm}{m}{n}
\rput(11.24,-2.255){$Y_C$}
\pspolygon[linewidth=0.04](10.388567,-2.3222651)(10.694077,-2.3222651)(10.541322,-2.532453)
\psline[linewidth=0.04cm](10.541322,-2.5124352)(10.541322,-2.7226233)
\psline[linewidth=0.03cm,arrowsize=0.05291667cm 2.0,arrowlength=1.4,arrowinset=0.4]{->}(7.2,0.24)(10.14,-2.54)
\psline[linewidth=0.03cm,arrowsize=0.05291667cm 2.0,arrowlength=1.4,arrowinset=0.4]{->}(7.18,-0.94)(10.1,-2.56)
\psellipse[linewidth=0.03,linestyle=dashed,dash=0.16cm 0.16cm,dimen=outer](8.91,-0.32)(0.15,1.84)
\psline[linewidth=0.03,arrowsize=0.05291667cm 2.0,arrowlength=1.4,arrowinset=0.4]{->}(6.64,2.68)(5.88,2.68)(5.9,0.92)
\usefont{T1}{ptm}{m}{n}
\rput(13.75,1.685){\small $L_1(u_A, v_A)+L_4(u_B,v_B)$}
\usefont{T1}{ptm}{m}{n}
\rput(13.71,2.425){$m=1$}
\psframe[linewidth=0.04,dimen=outer](15.6,2.06)(11.94,1.24)
\usefont{T1}{ptm}{m}{n}
\rput(13.75,-0.335){\small $L_2(u_A, v_A)+L_5(u_B,v_B)$}
\psframe[linewidth=0.04,dimen=outer](15.6,0.04)(11.94,-0.74)
\usefont{T1}{ptm}{m}{n}
\rput(13.73,-2.555){\small $L_3(u_A, v_A)+L_6(u_B,v_B)$}
\psframe[linewidth=0.04,dimen=outer](15.58,-2.14)(11.92,-2.98)
\usefont{T1}{ptm}{m}{n}
\rput(0.93,1.265){$m=1$}
\usefont{T1}{ptm}{m}{n}
\rput(2.67,1.285){$m=2$}
\usefont{T1}{ptm}{m}{n}
\rput(4.41,1.305){$m=3$}
\usefont{T1}{ptm}{m}{n}
\rput(17.84,1.705){\small $L_7(w_A, \psi_A)+L_{10}(u_C,v_C)$}
\usefont{T1}{ptm}{m}{n}
\rput(17.69,2.445){$m=2$}
\psframe[linewidth=0.04,dimen=outer](19.78,2.08)(15.92,1.28)
\psframe[linewidth=0.04,dimen=outer](19.82,0.06)(16.02,-0.72)
\psframe[linewidth=0.04,dimen=outer](19.84,-2.12)(16.0,-2.94)
\usefont{T1}{ptm}{m}{n}
\rput(17.86,-0.315){\small $L_8(w_A, \psi_A)+L_{11}(u_C,v_C)$}
\usefont{T1}{ptm}{m}{n}
\rput(17.86,-2.535){\small $L_9(w_A, \psi_A)+L_{12}(u_C,v_C)$}
\usefont{T1}{ptm}{m}{n}
\rput(22.11,1.665){\small $L_{13}(w_B, \psi_B)+L_{16}(w_C,\psi_C)$}
\usefont{T1}{ptm}{m}{n}
\rput(21.81,2.445){$m=3$}
\psframe[linewidth=0.04,dimen=outer](24.2,2.08)(20.04,1.28)
\psframe[linewidth=0.04,dimen=outer](24.16,0.06)(20.04,-0.7)
\psframe[linewidth=0.04,dimen=outer](24.22,-2.12)(20.02,-2.92)
\usefont{T1}{ptm}{m}{n}
\rput(22.07,-0.255){\small $L_{14}(w_B, \psi_B)+L_{17}(w_C,\psi_C)$}
\usefont{T1}{ptm}{m}{n}
\rput(22.11,-2.475){\small $L_{15}(w_B, \psi_B)+L_{18}(w_C,\psi_C)$}
\end{pspicture} 
}
\centering
\caption{Optimal Scheme for a System with $M=2$ and $K=3$, The First Phase}
\label{fig:feedback_24_Ph1}
\end{figure*}
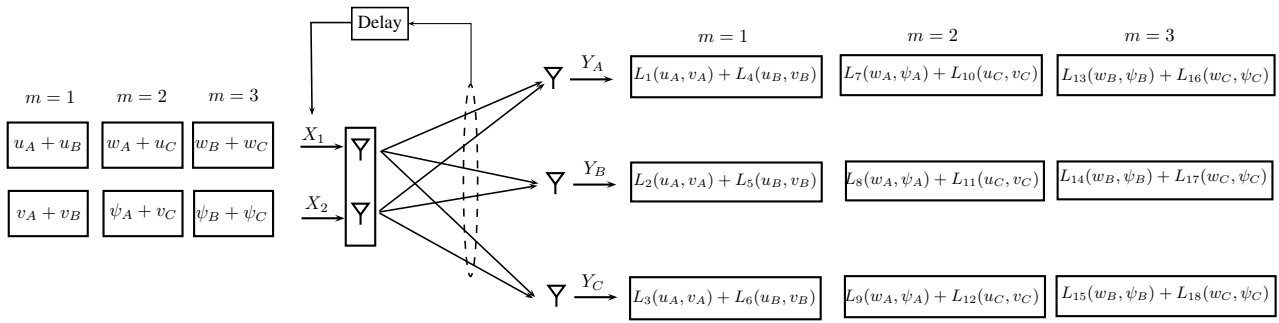

Let $u_r$, $v_r$, $w_r$, and $\psi_r$ be four symbols for receiver $r$, $r=A, B, C$. In the first time slot, which is dedicated to receivers $A$ and $B$, the transmitter sends random linear combinations of  four symbols $u_A$ and  $v_A$,  $u_B$,  and $v_B$.  Refer to Fig.~\ref{fig:feedback_24_Ph1} to see the details of particular  realizations for the linear combinations.  Receiver $A$ receives a linear combination of all four symbols denoted by $L_1(u_A, v_A)+L_4(u_B,v_B)$.    Receivers $B$ and $C$ also receive linear combinations of all four symbols denoted by $L_2(u_A, v_A)+L_5(u_B,v_B)$ and   $L_3(u_A, v_A)+L_6(u_B,v_B)$, respectively.  In the second time slot, which is dedicated to receivers $A$ and $C$, the transmitter sends random linear combinations of  four symbols $w_A$ and  $\psi_A$,  $u_C$,  and $v_C$. In the third time slot, which is dedicated to receivers $B$ and $C$, the transmitter sends random linear combinations of  four symbols $w_B$, $\psi_B$,  $w_C$,  and $\psi_C$.

By referring to Fig.~\ref{fig:feedback_24_Ph1}, it is easy to see that for each receiver to solve for all four desired symbols,  it is enough that
\begin{itemize}
\item receiver $A$ has $L_2(u_A,v_A)$, $L_4(u_B,v_B)$,  $L_9(w_A,\psi_A)$, and $L_{10}(u_C,v_C)$.
\item receiver $B$ has $L_2(u_A,v_A)$, $L_4(u_B,v_B)$,  $L_{15}(w_B,\psi_B)$, and $L_{17}(w_C,\psi_C)$.
\item receiver $C$ has  $L_9(w_A,\psi_A)$, $L_{10}(u_C,v_C)$, $L_{15}(w_B,\psi_B)$, and $L_{17}(w_C,\psi_C)$.
\end{itemize}
Therefore, the transmitter needs to deliver
\begin{itemize}
\item  $L_2(u_A,v_A)$ and $L_4(u_B,v_B)$ to both receivers $A$ and  $B$.
\item  $L_9(w_A,\psi_A)$ and  $L_{10}(u_C,v_C)$ to both receivers $A$ and  $C$.
\item  $L_{15}(w_B,\psi_B)$ and $L_{17}(w_C,\psi_C)$ to both receivers $B$ and $C$.
\end{itemize}

Therefore, we have 6 order--2 symbols as
\begin{align}
u_{AB}&=L_2(u_A,v_A), \qquad v_{AB}= L_4(u_B,v_B),\\
u_{AC}&=L_9(w_A,\psi_A), \qquad  v_{AC}=L_{10}(u_C,v_C),\\
u_{BC}&=L_{15}(w_B,\psi_B), \qquad v_{BC}=L_{17}(w_C,\psi_C).
\end{align}
Therefore, the transmitter needs $\frac{6}{{\DoF}_2^*(2,3)}$ more time--slots to deliver these 6 order--two symbols. Thus, we have
\begin{align}\label{DoF_1(2,3)_alt}
{\DoF}_1(2,3)=\frac{ 12 }{  3 +  \frac{6}{{\DoF}^*_{2}(2,3)}}=\frac{3}{2},
\end{align}
where we used Theorem~\ref{thm:squre} to set  ${\DoF}^*_{2}(2,3)=\frac{6}{5}$. Note the outer--bound in Theorem~\ref{thm:outer--bound--Gen} yields ${\DoF}^*_1(2,3) \leq \frac{3}{2}$, and therefore, this algorithm meets the outer--bound.  This result shows that the scheme of Section~\ref{Sec:non-square:original} is in general suboptimal.

\section{Connections with the Packet Erasure Broadcast Channel}
\label{sec:bec}

The schemes we proposed in this paper are inspired by schemes designed for the packet erasure broadcast channel, where each receiver observes the same transmitted packet but with a probability of erasure, and acknowledgement feedback is received by the transmitter from both receivers. Here, the delayed CSI that is fed back to the transmitter is the {\em erasure states} of the previous transmissions.

 The goal of these  packet erasure broadcast schemes is to exploit the fact that a packet intended for a receiver may be erased at that receiver but received at other receivers. These overheard packets become side information that can be exploited later. The basic scheme, initially proposed by~\cite{Jolfaei_XOR} for unicast setting, and then by~\cite{Larsson_ARQ} for multicasting setting, in the two--receiver case, works as follows. The transmitter sends packets intended for each receiver separately. If a packet is received by the intended receiver, then no extra effort is needed for that packet. But if a packet is received by the non-intended receiver, and not received by intended receiver, that receiver keeps that packet for later coding opportunity. Let us say packet $x_A$ intended for receiver $A$ is received by receiver $B$,  and packet $x_B$ intended for receiver $B$ is received by receiver $A$. In this case, the transmitter sends $(x_A\ \textrm{XOR} \ x_B)$. Then if receiver $A$ receives it, it can recover $x_A$ by subtracting $x_B$, and if receiver $B$ receives it, it can recover $x_B$ by subtracting $x_A$. In~\cite{Georgiadis_PEC}, the outer-bound of~\cite{Ozarow_BC_feedback} is used to show that the scheme of~\cite{Larsson_ARQ} is optimal. In~\cite{Wang_erasure_2_J, Tassiulas_erasure_K}, this two-receiver scheme is extended to more than two receivers, when all receivers have identical erasure probability. The scheme we proposed in this paper for the MIMO broadcast channel can be viewed as the counterpart to this scheme for the packet erasure broadcast channel.

\section{Follow-up Results}

\label{sec:followup}
 After the conference version of this paper has
appeared in~\cite{Maddah-Tse-Allerton}, the problem of exploiting
outdated CSIT in networks  have been investigated in several pieces
of work. In~\cite{Jafar_Retrospective}, it is shown that for
three-user interference channels and two--user X channels, outdated
CSIT can be used to achieve ${\DoF}$ more than one. In~\cite{GhasemiX1}, for two--user X channels, the result of~\cite{Jafar_Retrospective} has been improved and for three--user case, an achievable ${\DoF}$ has been proposed. In~\cite{Abdoli_K_Int}, an achievable  ${\DoF}$ for $K$--user single--antenna interference channels has been derived.  
In~\cite{ Abdoli3BC, Vaze_DCSIT_MIMO_BC,Ghasemi_DCSIT_Int},
the ${\DoF}$ regions of two-user and three--user MIMO broadcasts channels and
two-user MIMO interference channels with delayed CSIT are
studied.
 In~\cite{Xu_DCSIT}, the load of feedback to
implement the proposed scheme is evaluated. It is shown that for a
wide and practical range of channel parameters, the scheme of this
paper outperforms zero--forcing precoding and also single--user
transmission.

\section{Conclusions}
\label{sec:concl}

From the point of view of the role of feedback in information theory, this work provides yet another example that feedback can be useful in increasing the capacity of {\em multiuser} channels, even when the channels are memoryless. This is in contrast to Shannon's pessimistic result that feedback does not increase the capacity of memoryless {\em point-to-point} channels \cite{Shannon_zero_error}. In the specific context of broadcast channels, Ozarow \cite{Ozarow_BC_feedback} has in fact already  shown that feedback can increase the capacity of Gaussian scalar non-fading broadcast channels. However, the nature of the gain is unclear, as it was shown numerically. Moreover, the gain is quite limited. We argue that the MIMO fading broadcast channel considered in this paper provides a much more interesting example of the role of feedback. The nature of the gain is very clear. In contrast to the Gaussian scalar non-fading broadcast channel, the main uncertainty from the point of the view of the transmitter is the channel direction rather than the additive noise, particularly in the high $\SNR$ regime.  This means that although the MIMO channel has intrinsically multiple degrees of freedom, the transmitter cannot segregate it into multiple orthogonal channels, one for each receiver. Hence, when transmitting information for one receiver, significant part of that information is overheard at other receivers. This overheard information becomes side information that can be exploited in future transmissions. The role of feedback is to provide the channel directions to the transmitter {\em after} the transmission to allow the transmitter to determine the side information that was received at the receivers. Overall, feedback leads to a much more efficient use of the intrinsic multiple degrees of freedom in the MIMO channel, yielding a multiplexing gain over the non-feedback case.

\appendices

\section{An Identity} \label{Appendix-A}

In this appendix, we prove that for any $j$,  $1 \leq  j \leq K$,
\begin{align}\label{eq:identity}
\frac{1}{ {K \choose j-1}}\sum_{i=1}^{K-j+1}\frac{ {K-i \choose j-1} }{i} =\sum_{i=j}^{K}\frac{1}{i}.
\end{align}
We define LHS of \eqref{eq:identity} as  $f(j)$,
\begin{align}
f(j) = \frac{1}{ {K \choose j-1}}\sum_{i=1}^{K-j+1}\frac{ {K-i \choose j-1} }{i}.
\end{align}
Then it is easy to see that $f(K)=\frac{1}{K}$. In what follows, we prove that for any $j$,  $ 1 \leq j \leq K-1$,
\begin{align*}
f(j)=\frac{1}{j}+f(j+1),
\end{align*}
which yields identity \eqref{eq:identity}.

We have
\begin{align*}
& f(j)-f(j+1)   \\
& =\frac{1}{ {K \choose j-1}}\sum_{i=1}^{K-j+1}\frac{ {K-i \choose j-1} }{i} - \frac{1}{ {K \choose j}}\sum_{i=1}^{K-j}\frac{ {K-i \choose j} }{i} \\
& = \frac{ (j-1)! (K-j)!}{ K!} \\
& \times \left\{ \sum_{i=1}^{K-j+1}\frac{(K-j+1) {K-i \choose j-1} }{i} - \sum_{i=1}^{K-j}\frac{j {K-i \choose j} }{i} \right\} \\
& = \frac{ (j-1)! (K-j)!}{ K!} 
\\
& \times \left\{ \sum_{i=1}^{K-j}\frac{ {K-i \choose j-1}  [(K-j+1)-(K-i-j+1) ]}{i}  +1\right\}  \\
& = \frac{ (j-1)! (K-j)!}{ K!} \sum_{i=1}^{K-j+1} {K-i \choose j-1}  \\
& = \frac{ (j-1)! (K-j)!}{ K!} \sum_{l=j-1}^{K-1} {l \choose j-1}  \\
& \overset{(a)}{=} \frac{ (j-1)! (K-j)!}{ K!} {K \choose j}  \\
& = \frac{1}{j},
\end{align*}
where $(a)$ follows from the identity that
\begin{align}\label{eq:identity2}
\sum_{l=p}^{q} {l \choose p} = {q+1 \choose p+1}, \qquad 0 \leq p \leq q.
\end{align}
Equation \eqref{eq:identity2} can simply be proved by induction.


\bibliographystyle{IEEE}

\end{document}